\documentclass[aip, amsfonts, amssymb, amsmath , onecolumn, reprint, subscriptadress, showpacs]{revtex4}
\usepackage{amssymb}		 
\usepackage{amsmath, amsthm, amssymb,amsfonts}
\usepackage[english]{babel}
\usepackage{color, graphicx}
\usepackage{epstopdf} 
\usepackage{caption}
\usepackage{comment}
\usepackage{hyperref}

\newcommand\eTF{{\em e}TF}

\newcommand\Bv{{\bf B}}
\newcommand\bv{{\bf b}}

\newcommand\ev{{\bf e}}
\newcommand\uv{{\bf u}}

\newcommand\Pia{\boldsymbol{\Pi}_\alpha}
\newcommand\wPia{\widetilde{\boldsymbol{\Pi}}_\alpha}

\newcommand\Piaij{\Pi_{\alpha,ij}}
\newcommand\Piaik{\Pi_{\alpha,ik}}
\newcommand\Piajk{\Pi_{\alpha,jk}}
\newcommand\Piail{\Pi_{\alpha,il}}
\newcommand\Piajl{\Pi_{\alpha,jl}}

\newcommand\Piaxx{\Pi_{\alpha,xx}}
\newcommand\Piayy{\Pi_{\alpha,yy}}
\newcommand\Piazz{\Pi_{\alpha,zz}}

\newcommand\wPiaxx{\widetilde{\Pi}_{\alpha,xx}}
\newcommand\wPiayy{\widetilde{\Pi}_{\alpha,yy}}
\newcommand\wPiazz{\widetilde{\Pi}_{\alpha,zz}}

\newcommand\Pixx{\Pi_{xx}}
\newcommand\Piyy{\Pi_{yy}}
\newcommand\Pizz{\Pi_{zz}}

\newcommand\Pie{\boldsymbol{\Pi}_{\rm e}}
\newcommand\Piixx{\Pi_{{\rm i},xx}}
\newcommand\Piiyy{\Pi_{{\rm i},yy}}
\newcommand\Piizz{\Pi_{{\rm i},zz}}
\newcommand\Piexx{\Pi_{{\rm e},xx}}
\newcommand\Pieyy{\Pi_{{\rm e},yy}}
\newcommand\Piezz{\Pi_{{\rm e},zz}}
\newcommand\pperp{p_\perp}

\newcommand\piperp{p_{{\rm i},\perp}}
\newcommand\pipara{p_{{\rm i},\|}}
\newcommand\peperp{p_{{\rm e},\perp}}
\newcommand\pepara{p_{{\rm e},\|}}
\newcommand\Tiperp{T_{{\rm i},\perp}}
\newcommand\Tipara{T_{{\rm i},\|}}
\newcommand\Teperp{T_{{\rm e},\perp}}
\newcommand\Tepara{T_{{\rm e},\|}}
\newcommand\piperpz{p_{{\rm i},\perp0}}
\newcommand\piparaz{p_{{\rm i},\|0}}
\newcommand\peperpz{p_{{\rm e},\perp0}}
\newcommand\peparaz{p_{{\rm e},\|0}}
\newcommand\paperp{p_{\alpha,\perp}}
\newcommand\papara{p_{\alpha,\|}}
\newcommand\paperpz{p_{\alpha,\perp0}}

\newcommand\Tiperpz{T_{{\rm i},\perp0}}
\newcommand\Tiparaz{T_{{\rm i},\|0}}
\newcommand\Teperpz{T_{{\rm e},\perp0}}

\newcommand\Qaijk{Q_{\alpha,ijk}}

\newcommand\eijk{\epsilon_{ijk}}
\newcommand\eilm{\epsilon_{ilm}}
\newcommand\ejlm{\epsilon_{jlm}}

\newcommand\ma{m_\alpha}
\newcommand\qa{q_\alpha}
\newcommand\sigmaa{\sigma_\alpha}
\newcommand\Omegaa{\Omega_\alpha}
\newcommand\Omegav{\boldsymbol{\Omega}}
\newcommand\Bmod{|{\bf B}|}
\newcommand\uai{u_{\alpha,i}}
\newcommand\uaj{u_{\alpha,j}}
\newcommand\uak{u_{\alpha,k}}

\newcommand\duaydx{\frac{du_{\alpha,y}}{dx}}

\newcommand\uay{u_{\alpha,y}}

\newcommand\uiy{u_{{\rm i},y}}

\newcommand\duiydx{\frac{du_{{\rm i},y}}{dx}}

\newcommand\Aperp{\mathcal{A}_\perp}
\newcommand\Aaperp{\mathcal{A}_{\alpha,\perp}}

\newcommand\ai{a_{\rm i}}
\newcommand\azi{a_{0{\rm i}}}
\newcommand\FF{\mathcal{F}}
\newcommand\GG{\mathcal{G}}
\newcommand\HH{\mathcal{H}}
\newcommand\CC{\mathcal{C}}
\newcommand\wgam{\widetilde{\gamma}}
\newcommand\giperp{\gamma_{{\rm i},\perp}}
\newcommand\geperp{\gamma_{{\rm e},\perp}}
\newcommand\gipara{\gamma_{{\rm i},\|}}
\newcommand\gepara{\gamma_{{\rm e},\|}}
\newcommand\betiprp{\beta_{{\rm i},\perp}}

\newcommand\betperp{\beta_\perp}
\newcommand\betprpz{\beta_{\perp0}}
\newcommand\betaprpz{\beta_{\alpha,\perp0}}
\newcommand\betiprpz{\beta_{{\rm i},\perp0}}
\newcommand\beteprpz{\beta_{{\rm e},\perp0}}
\newcommand\wbetiprpz{\widetilde{\beta}_{{\rm i},\perp0}}
\newcommand\wbeteprpz{\widetilde{\beta}_{{\rm e},\perp0}}
\newcommand\wbetaprpz{\widetilde{\beta}_{\alpha,\perp0}}
\newcommand\wFFg{\widetilde{\mathcal{F}}_{\wgam}}
\newcommand\tauprp{\tau_{_\perp}}

\begin{document}

\title{Pressure tensor in the presence of velocity shear: stationary solutions and self-consistent equilibria}
\author{S. S. Cerri}\email{silvio.sergio.cerri@ipp.mpg.de}\affiliation{Max-Planck-Institut f\"ur Plasmaphysik, Boltzmannstr. 2, D-85748 Garching, Germany}
\author{F. Pegoraro}\affiliation{Physics Department ``E. Fermi'', University of Pisa, Largo B. Pontecorvo 3, 56127 Pisa, Italy}
\author{F. Califano}\affiliation{Physics Department ``E. Fermi'', University of Pisa, Largo B. Pontecorvo 3, 56127 Pisa, Italy}\affiliation{Max-Planck/Princeton Center for Plasma Physics}
\author{D. Del Sarto}\affiliation{Institut Jean Lamour, UMR 7198 CNRS - Universit\'e de Lorraine, BP 239 F-54506 Vandoeuvre les Nancy, France}
\author{F. Jenko}\affiliation{Max-Planck-Institut f\"ur Plasmaphysik, Boltzmannstr. 2, D-85748 Garching, Germany}\affiliation{Department of Physics and Astronomy, University of California, Los Angeles, CA 90095, USA}\affiliation{Max-Planck/Princeton Center for Plasma Physics}

\begin{abstract}
Observations and numerical simulations of laboratory and space plasmas in almost collisionless regimes reveal anisotropic and non-gyrotropic  particle distribution functions. We investigate how such states can persist in the presence of a sheared flow. We focus our attention on the pressure tensor equation in a magnetized plasma and derive analytical self-consistent plasma equilibria which exhibit a novel asymmetry with respect to the magnetic field direction. These results are relevant for investigating, within fluid models that retain the full pressure tensor dynamics, plasma configurations where a background shear flow is present.
\end{abstract}


\maketitle

\section{Introduction}\label{sec:intro}

Sheared flows are frequently observed in space and laboratory plasmas. They are an important source of free energy and can drive various instabilities, such as the Kelvin-Helmholtz instability~\cite{NaganoPSS1979,HubaGRL1996,MiuraPOP1997,FaganelloPRL2008a,CalifanoNPG2009,FaganelloNJP2009,NakamuraPOP2010,NakamuraJGR2011,PalermoJGR2011a,HenriPOP2013} (KHI) or the magnetorotational instability~\cite{VelikhovJETP1959,ChandrasekharPNAS1960,SharmaAPJ2006,FerraroAPJ2007,RenPPCF2011,RiquelmeAPJ2012,ShirakawaPOP2014} (MRI). The KHI, on the one hand, leads to the formation of fully developed large scale vortices, eventually ending in a turbulent state where energy is efficiently transferred to small scales. 
In this context, a relevant example is given by the development of the KHI observed at the flanks of the Earth's magnetosphere~\cite{HasegawaNATURE2004} driven by the velocity shear between the solar wind (SW) and the magnetosphere (MS) plasma, and in general observed at other planetary magnetospheres. On the other hand, the MRI is considered to be a main driver of turbulence (and turbulent transport of angular momentum) in accretion disks around astrophysical objects, such as stars and black holes. In addition, small-scale sheared flows can emerge from turbulent states and lead to kinetic anisotropy effects, as seen from SW data and simulations~\cite{ServidioPRL2012,ServidioAPJL2014,ValentiniPOP2014}.

The standard magnetohydrodynamic (MHD) approach to the study of shear flow configurations is justified when the scale length of the sheared flow is much larger than the typical ion microscales, i.e. when $d_{\rm i}, \rho_{\rm i}\ll L$. However, in the case of the interaction of the SW with the MSP, satellite observations show that the typical scale length of the sheared flow is roughly comparable to the ion gyroradius and/or the skin depth, i.e. $d_{\rm i}\sim\rho_{\rm i}\lesssim L$ ($\beta\sim1$). In general, a common assumption in the framework of MHD modeling is to consider an isotropic pressure tensor even in the presence of a background magnetic field. However, a relatively simple step to avoid such extreme simplification is to adopt the Chew-Goldberger-Low (CGL) approximation~\cite{CGL1956}, where the pressure tensor is gyrotropic, i.e. the pressure can be different along the magnetic field and perpendicular to it. 
In this approach, three main features of the system appear to be relevant: (i) the pressure is isotropic in the plane perpendicular to the magnetic field, i.e. {\em gyrotropy}, (ii) the system is {\em symmetric} with respect to the relative orientation of the magnetic field $\Bv$ and the fluid vorticity $\Omegav_\uv\equiv\nabla\times\uv$, i.e. with respect to the sign of $\Omegav_\uv\cdot\Bv$, and (iii) the equilibrium profiles are not dependent on the velocity shear. These three points are substantially modified when the pressure tensor equation is retained in the fluid hierarchy or when kinetic models are adopted. Also retaining first-order finite Larmor radius (FLR) corrections of the ions within a two-fluid (TF) model, the so called {\em extended} two-fluid (\eTF) model~\cite{CerriPOP2013}, was recently shown to substantially modify the previous picture.

In this work, we investigate the role of retaining the full pressure tensor equation, still in the framework of a fluid model. In order to simplify the picture for the sake of clarity, we will consider a configuration in which the inhomogeneity direction, the flow direction and the magnetic field are orthogonal to each other. The main result of our approach is to prove that, in the presence of a shear flow, an additional anisotropy in the perpendicular plane ({\em agyrotropy}) and an {\em asymmetry} with respect to the sign of $\Omegav_\uv\cdot\Bv$ arise even at the level of the equilibrium configurations, which depend also on the shear strength.
{Indeed, a sheared flow can induce dynamical anisotropization of an initial isotropic pressure configuration, together with an asymmetry with respect to the sign of $\Omegav_\uv\cdot\Bv$, when, for instance, one retains first-order FLR corrections~\cite{CerriPOP2013} or the full pressure tensor equation~\cite{DelSarto_TensPress}. Here, we focus our attention on the effect of the shear-induced anisotropization  at the level of a stationary state.  Such new features are intrinsic properties of the system and their relevance is related, in general, to the plasma regime under investigation. In particular, the deviation from an MHD/CGL model becomes not negligible when the shear length scale is comparable with the ion microscales ($d_{\rm i}, \rho_{\rm i}\lesssim L$), as is the case for the SW-MS interaction. Such deviations, important at the level of equilibrium configurations, can dramatically affect the study of one of the above mentioned instabilities already in the linear phase and possibly leading to very different nonlinear stages, when kinetic models are adopted~\cite{HenriPOP2013}. This highlights the importance of a correct modeling of the sheared flow equilibrium configuration, in which the possibly relevant ingredients, such as the pressure tensor, are retained. Moreover, as already anticipated, the approach presented here can give insights into non-gyrotropic proton distribution functions that are observed in SW data and Vlasov simulations~\cite{ServidioPRL2012,ServidioAPJL2014,ValentiniPOP2014}.

The remainder of this paper is organized as follows. In Sec. \ref{sec:Pi_solution}, we solve the stationary pressure tensor equation without heat fluxes and in the presence of a sheared flow, giving the solution in the form of traceless corrections to the CGL gyrotropic pressure tensor, discussing the emergence of the perpendicular anisotropy and of the $\Omegav_\uv\cdot\Bv$-asymmetry. In Sec. \ref{sec:FLRequilibria}, we consider the full non-gyrotropic ion pressure tensor within the equilibrium problem. Implicit and exact numerical solution for the equilibrium profiles are then given, along with possibly useful explicit analytical approximations, underlining again the role of the asymmetry and the perpendicular anisotropy. Finally, alternative approximated and exact analytical equilibria are given in Appendix \ref{app:FLRequilibria_altern}.

\section{Stationary solution of the pressure tensor equation}\label{sec:Pi_solution}

Within a fluid description of a plasma, the pressure tensor equation is given by~\cite{CerriPOP2013}
\begin{equation}\label{eq:Pi-eq_full}
\frac{\partial\Piaij}{\partial t} + \frac{\partial}{\partial x_k}\left(\Piaij\uak+\Qaijk\right) +
\Piaik\frac{\partial\uaj}{\partial x_k} + \Piajk\frac{\partial\uai}{\partial x_k} =
\frac{\qa}{\ma c}\left(\eilm\Piajl+\ejlm\Piail\right)B_m\,,
\end{equation}
where $\qa$ and $\ma$ are the charge and the mass of the species $\alpha$, respectively, $\Piaij$ is the ($ij$-component of the) pressure tensor, $\uak$ is the ($k$-component of the) fluid velocity, $\Qaijk$ is the ($ijk$-component of the) heat flux tensor, $\eijk$ is the Levi-Civita symbol and $B_m$ is the ($m$-component of the) magnetic field. Now, we look for stationary solutions of Eq.~(\ref{eq:Pi-eq_full}) in the limit of no heat fluxes, i.e. $\partial\Piaij/\partial t=0$ and $\Qaijk=0$ $\forall$ $i,j,k$. Under these assumptions, Eq.~(\ref{eq:Pi-eq_full}) reduces to
\begin{equation}\label{eq:Pi-eq_red-1}
\frac{\partial}{\partial x_k}\left(\Piaij\uak\right) +
\Piaik\frac{\partial\uaj}{\partial x_k} + \Piajk\frac{\partial\uai}{\partial x_k} =
\sigmaa\Omegaa\left(\eilm\Piajl+\ejlm\Piail\right)b_m\,,
\end{equation}
where we have introduced the sign of the charge $\sigmaa\equiv{\rm sign}(\qa)$ and the cyclotron frequency $\Omegaa\equiv|\qa|\Bmod/\ma c$ of the species $\alpha$, respectively, and the magnetic field versor $b_m\equiv B_m/\Bmod$.

It is a well known result that, within a finite Larmor radius (FLR) expansion of Eq.~(\ref{eq:Pi-eq_red-1}), the zero-order solution is given by the CGL pressure tensor~\cite{CGL1956,CerriPOP2013},
\begin{equation}\label{eq:CGL_solution}
 \Pia^{(0)}= \paperp\boldsymbol{\tau} + \papara\bv\bv
\end{equation}
where $\boldsymbol{\tau}\equiv{\bf I}-\bv\bv$ is the projector onto the plane perpendicular to the magnetic field ($\bv\equiv\Bv/\Bmod$),  $\papara$ and $\paperp$ are the pressure parallel and perpendicular to $\Bv$, respectively. The tensor $\Pia^{(0)}$ represents the kernel of the operator on the right hand side of the pressure tensor equation, i.e. it is an approximated solution of the equation when the left hand side is negligible. 

We now consider a velocity shear configuration such that the inhomogeneity direction, the flow direction and the magnetic field direction form a right-handed basis, e.g. $\uv=u_y(x)\ev_y$ and $\Bv=B_z(x)\ev_z$. Note that this configuration, despite its apparent simplicity, is actually commonly used in various areas, e.g. for studying the KHI or the MRI. Without loss of generality we can write the full pressure tensor as $\Pia=\Pia^{(0)}+\wPia$, in which $\wPia$ represents a traceless correction to the gyrotropic pressure tensor $\Pia^{(0)}$. In fact, remaining within a fluid framework, one has unambiguous definitions of $\paperp\equiv\frac{1}{2}\Pia:\boldsymbol{\tau}$ and $\papara\equiv\Pia:\bv\bv$, so the identity $\wPia:{\bf I}=0$ holds~\cite{RosenbluthPOF1965,MacmahonPOF1965,RamosPOP2005,GoswamiPOP2005,PassotPOP2007,SulemJPP2014}.  The same conclusion can be directly derived  from the Vlasov equation~\cite{SchekochihinMNRAS2010}. From the definition of $\papara$ one obtains also $\wPia:\bv\bv=0$, so in our configuration $\wPiazz=0$ and the perpendicular components of the pressure tensor read
\begin{equation}\label{eq:traceless-condition}
 \Piaxx = \paperp + \wPiaxx\quad,\quad
 \Piayy = \paperp + \wPiayy\quad{\rm with}\quad
 \wPiayy = - \wPiaxx\,.
\end{equation}
Inserting the above expressions into Eq.~(\ref{eq:Pi-eq_red-1}) gives the following non-gyrotropic diagonal pressure tensor:
\begin{equation}\label{eq:Pi_solution}
 \left\{\begin{array}{c}
   \Piazz = \papara\\
   \\
   \Piaxx = \left(1-\frac{a_\alpha(x)}{1+a_\alpha(x)}\right)\paperp\\
   \\
   \Piayy = \left(1+\frac{a_\alpha(x)}{1+a_\alpha(x)}\right)\paperp\\
 \end{array}\right.
\end{equation}
with
\begin{equation}\label{eq:a_def}
 a_\alpha(x) \equiv \frac{1}{2}\frac{s_3\,\sigmaa}{\Omegaa}\duaydx\,,
\end{equation}
where $s_3\equiv{\rm sign}(b_3)$ is the relative orientation of the magnetic field and the $z$-axis (see Ref.~\cite{CerriPOP2013}). The positivity condition on the diagonal pressure terms in Eq.~(\ref{eq:Pi_solution}) gives
\begin{equation}\label{eq:a_positiv-cond}
 a_\alpha(x) \geq -\frac{1}{2}\,.
\end{equation}
It is interesting to note that here an asymmetry with respect to the sign of $a_\alpha(x)$ appears. In fact, not all the value of the shear strength are allowed when $a_\alpha$ is negative, while in principle there is no limitation when it is positive. The sign of $a_\alpha$ depends essentially on two physical factors: the species, through $\sigmaa$, and the relative orientation of the magnetic field $\Bv$ and the fluid vorticity $\Omegav_\uv$, through the sign of $s_3(d\uay/dx)$. For instance, if one considers ions ($\sigma_{\rm i}=+1$) and a background magnetic field oriented in the positive direction of the $z$-axis ($s_3=+1$), from Eq.~(\ref{eq:a_positiv-cond}) we find that there is no limitation in the velocity shear of $\uv_{\rm i}$ if the vorticity is aligned with the magnetic field ($\Omegav_\uv\cdot\Bv>0$), while the velocity shear is limited to be maximum equal (in absolute value) to the ion gyration frequency otherwise ($|\Omegav_\uv|\leq\Omega_{\rm i}$). We note that this condition on the shear strength limitation asymmetry is indeed in accordance with Ref.~\cite{DelSarto_TensPress}, where it is found that shear configurations of the type considered here become dynamically unstable when $\Omega'/\Omega_{\rm i}\equiv1+(\partial_x u_{{\rm i},y})/\Omega_{\rm i}<0$, which corresponds exactly to the condition $a_{\rm i}(x)<-1/2$ in our notation (in the case of Ref.~\cite{DelSarto_TensPress}, however, the non-stationary pressure tensor equation was considered). 

The solution in Eq.~(\ref{eq:Pi_solution}) introduces an additional anisotropy in the plane perpendicular to the magnetic field:
\begin{equation}\label{eq:Aperp_def}
 \Aaperp \equiv \frac{\left|\Piaxx-\Piayy\right|}{\paperp} =
 2\left|\frac{a_\alpha(x)}{1+a_\alpha(x)}\right|\,,
\end{equation}
which gives the maximum value of the anisotropy ($\Aperp^{\rm(max)}=2$) for $a=-1/2$ and for $a\to\infty$. A sketch of $\Pixx/\pperp$, $\Piyy/\pperp$ and $\Aperp$ versus $a$ is given in Fig.~\ref{fig:fig0}.  The asymmetry is thus found also in the perpendicular anisotropy, which is here entirely due to the velocity shear.
\begin{figure}[!h]
  \begin{minipage}[!h]{0.49\textwidth}
  \flushleft\includegraphics[width=1.0\textwidth]{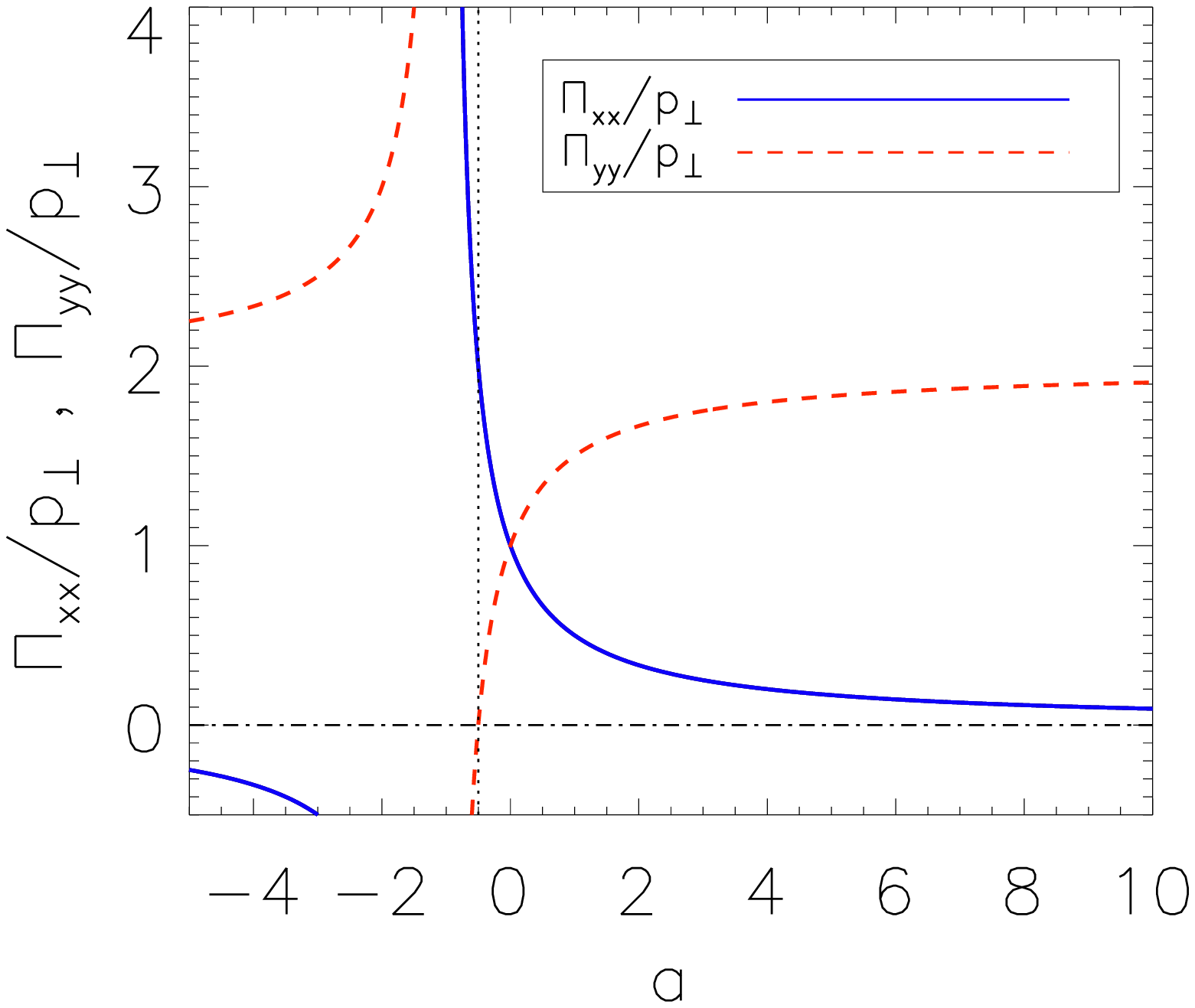}
  \end{minipage}
  \begin{minipage}[!h]{0.49\textwidth}
  \flushleft\includegraphics[width=1.0\textwidth]{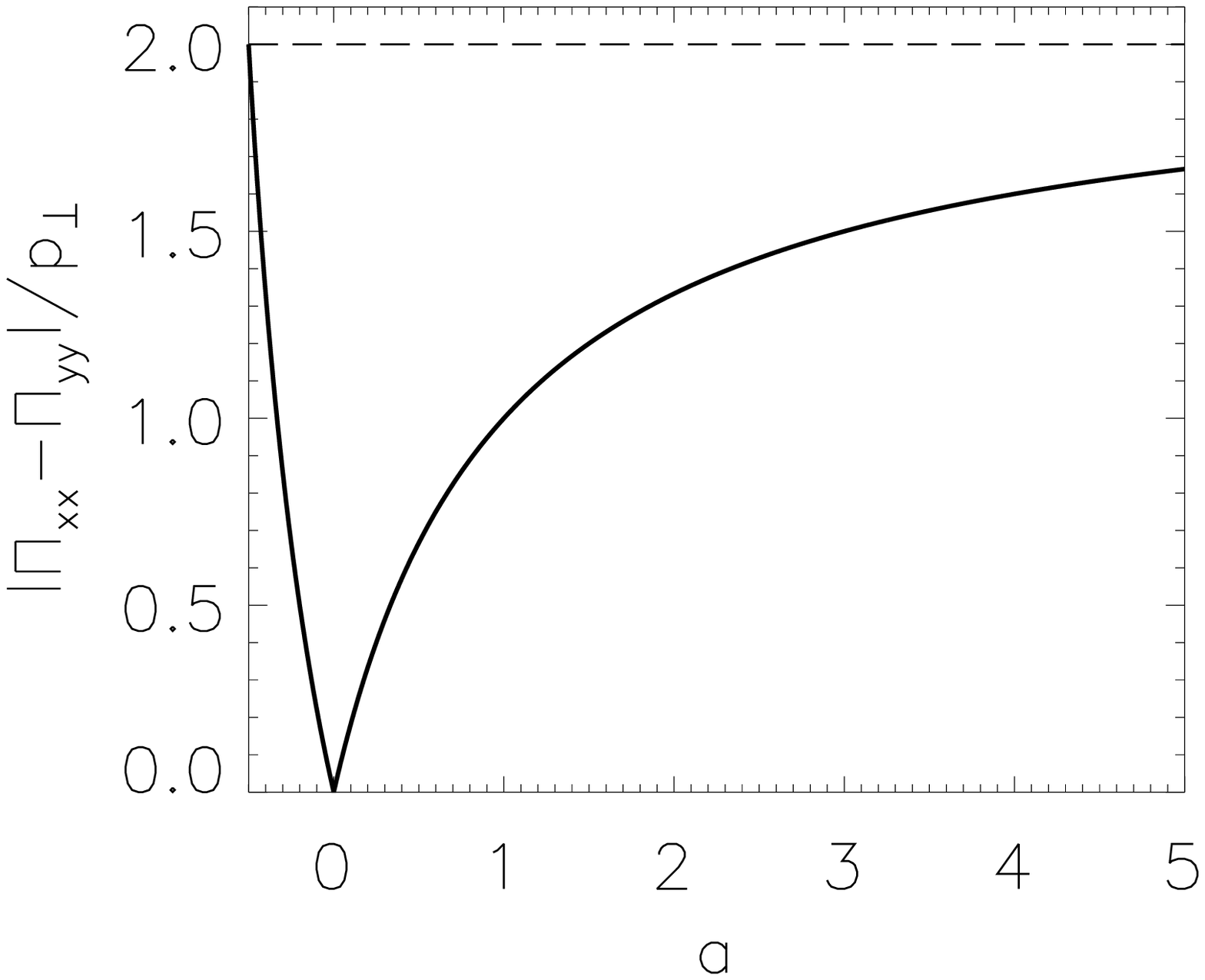}
  \end{minipage}
\caption{Left: plot of the solutions $\Pixx/\pperp$ (blue solid line) and $\Piyy/\pperp$ (red dashed line). Right: consequent perpendicular anisotropy $\Aperp$ (solid line) versus the shear parameter $a$.}
\label{fig:fig0}
\end{figure} 
\\
For small $|a|\ll1$, the corrections to the gyrotropic CGL pressure tensor are small and, if we expand the solution in Eq.~(\ref{eq:Pi_solution}) to the leading order in $a$, the first-order FLR solution is found~\cite{CerriPOP2013,RosenbluthPOF1965,MacmahonPOF1965,RamosPOP2005,GoswamiPOP2005,PassotPOP2007,SulemJPP2014}. However, in this limit, the asymmetry with respect to the sign of $a$ in Eq.~(\ref{eq:a_positiv-cond}) is lost.

\section{Equilibria with the complete pressure tensor}\label{sec:FLRequilibria}

We now want to use the stationary solution of $\Pixx$ in Eq.~(\ref{eq:Pi_solution}) for solving the equilibrium condition~\cite{CerriPOP2013}
\begin{equation}\label{eq:equil_cond}
 \frac{d}{dx}\left[\sum_\alpha\Piaxx + \frac{B^2}{8\pi}\right] = 
 \frac{d}{dx}\left[\Piixx + \peperp + \frac{B^2}{8\pi}\right] = 0\,,
\end{equation}
where here we are considering only the ions full pressure tensor, while the electrons are gyrotropic, i.e. $\Pie=\peperp\boldsymbol{\tau}+\pepara\bv\bv$. Note that in our configuration, since there are no parallel gradients, the parallel balance is automatically satisfied~\cite{PassotPOP2006}.
We define the gyrotropic pressure and the magnetic field profiles as
\begin{equation}\label{eq:MHD-FLR_functions}
\left\{\begin{array}{c}
 \piperp(x) = \piperpz\FF(x)f(x)\\
 \\
 \peperp(x) = \peperpz\GG(x)g(x)\\
 \\
 B^2(x) = B_0^2\HH(x)h(x)
\end{array}\right.
\end{equation}
where $\piperpz$ and $\peperpz$ are positive constants, $\FF(x)$, $\GG(x)$ and $\HH(x)$ correspond to the MHD equilibrium and $f(x)$, $g(x)$ and $h(x)$ are the corrections to the relative MHD profile. The equilibrium condition, Eq.~(\ref{eq:equil_cond}), can then be conveniently rewritten in dimensionless form as
\begin{equation}\label{eq:equil_cond_dimensionless}
 \wbetiprpz\left[1-\frac{\ai(x)}{1+\ai(x)}\right]\FF(x)f(x)+\wbeteprpz\GG(x)g(x)+\frac{\HH(x)h(x)}{1+\betprpz} - 1 = 0\,,
\end{equation}
where we have fixed the constant of integration to be $B_0^2/8\pi + \piperpz + \peperpz$ and the quantities $\wbetaprpz\equiv\betaprpz/(1+\betprpz)$, $\betprpz=\sum_\alpha\betaprpz$ and $\betaprpz\equiv8\pi\paperpz/B_0^2$ are introduced.
The equation can be further simplified. First, we note that the MHD equilibrium functions are related by the quasi-neutrality requirement and the MHD equilibrium conditions, i.e. quasi-neutrality gives~\cite{CerriPOP2013}
\begin{equation}\label{eq:MHD_quasi-neutr}
 \GG(x) = \left[\FF(x)\right]^{1+\wgam}\,,
\end{equation}
where $\wgam\equiv(\geperp/\giperp)-1$ and a polytropic relation between the pressure and the density is assumed~\cite{Note1}, while the MHD force balance condition gives~\cite{CerriPOP2013}
\begin{equation}\label{eq:MHD_balance}
 \HH(x) = 1 + \betprpz - \left[\betiprpz + \beteprpz\left(\FF(x)\right)^{\wgam}\right]\FF(x)\,.
\end{equation}
Then, we require again quasi-neutrality for our modified equilibrium, i.e.
\begin{equation}\label{eq:FLR_quasi-neutr}
 \GG(x)g(x) = \left[\FF(x)f(x)\right]^{1+\wgam}
 \quad{\rm or,\ equivalently,\ }\quad g(x) = \left[f(x)\right]^{1+\wgam}\,,
\end{equation}
where the equivalence is because of condition (\ref{eq:MHD_quasi-neutr}). We then require that the perpendicular plasma beta $\betiprp(x)$ remains unchanged with respect to the MHD equilibrium (see Appendix \ref{app:FLRequilibria_altern} for alternative requests), which then leads to the condition
\begin{equation}\label{eq:FLR_betaiperp_cond}
 h(x) = f(x)\,.
\end{equation}
If $\wgam=0$ holds, this condition is equivalent to the requirement that the total perpendicular plasma beta $\betperp(x)$ remains unchanged. Thus, using the previous relations, the equilibrium condition (\ref{eq:equil_cond_dimensionless}) reads
\begin{equation}\label{eq:FLR_equil_cond}
 \left[ 1 - 
 \wbetiprpz\left(\frac{\ai(x)}{1+\ai(x)}\right)\FF(x) 
 + \wbeteprpz\left(\FF(x)\right)^{1+\wgam}\left(f^{\wgam}(x)-1\right)\right]f(x) - 1 = 0\,,
\end{equation}
which is intrinsically nonlinear in $f$, not only because of the parameter $\wgam$, but also because $\ai(x)$ itself contains the magnetic field profile which must be derived self-consistently from the equilibrium, i.e.
\begin{equation}\label{eq:a_fullB}
 \ai(x) = \frac{1}{\sqrt{\HH(x)h(x)}}\frac{s_3\,m_i\,c}{2\,e\,B_0}\duiydx 
 \equiv \frac{\azi(x)}{\sqrt{h(x)}}\,,
\end{equation}
where for convenience we have separated $\sqrt{h(x)}$ and $\azi(x)$, since the latter term does not depend on the self-consistent solution $h(x)$. This intrinsic nonlinearity is the reason that justifies our approach of extending an MHD/CGL equilibrium to the ``corresponding'' full pressure tensor equilibrium: in the MHD/CGL case, we do not have $\ai(x)$, so the equilibrium condition is easily solvable and we find  $\FF(x)$, $\GG(x)$ and $\HH(x)$. Then, we can compute $\azi(x)$ and give the solution in terms of it. Moreover, the MHD equilibria are the most commonly adopted for various simulation initialization, even in a kinetic framework~\cite{HenriPOP2013}, and thus using them as a starting point may be convenient.\\
In the following, when we give explicit examples, a velocity profile described by a hyperbolic tangent will be adopted:
\begin{equation}\label{eq:uiy_tanh}
\uiy(x) = u_0\tanh\left(\frac{x-x_0}{L_u}\right)\,,
\end{equation}
which is often used for the study of KHI~\cite{FaganelloPRL2008a,CalifanoNPG2009,FaganelloNJP2009}, and all the quantities will be given in units of ions quantities ($m_i$, $e$, $\Omega_{\rm i}$, $d_{\rm i}$) and Alfv\`en velocity ($v_A$).

\subsection{Discussion of the fully self-consistent equilibria}\label{subsec:FLRequilibria_fully}

Let us consider the complete problem in which the $\ai(x)$ function is computed with the actual self-consistent magnetic field profile, Eq.~(\ref{eq:a_fullB}). For $\azi$ the positivity condition of the pressure, $\ai\geq-1/2$, reads
\begin{equation}\label{eq:azi_condition}
 \azi(x) \geq -\frac{\sqrt{h(x)}}{2}\qquad\forall\ x\,.
\end{equation}
Then, under the assumption of quasi-neutrality and the request that the perpendicular plasma beta $\betiprp(x)$ remains unchanged with respect to the MHD equilibrium, Eqs.~(\ref{eq:FLR_quasi-neutr})-(\ref{eq:FLR_betaiperp_cond}), the equilibrium condition (\ref{eq:FLR_equil_cond}) can be recast in the following form:
\[
\wbeteprpz\FF(x)^{1+\wgam}f(x)^{3/2+\wgam} +
 \left[1-\wbeteprpz\FF(x)^{1+\wgam}\right]f(x)^{3/2} +
 \wbeteprpz\FF(x)^{1+\wgam}\azi(x)f(x)^{1+\wgam}
\]
\begin{equation}\label{eq:EquilCond_fully-1}
 + \left[1-\wbetiprpz\FF(x)-\wbeteprpz\FF(x)^{1+\wgam}\right]\azi(x)f(x)
 - f(x)^{1/2} - \azi(x) = 0\,,
\end{equation}
which is absolutely non trivial, since the parameter $\wgam$ can change the order of the equation in a non obvious way. We thus restrict the problem to the case $\wgam=0$, i.e. to the case of an equal polytropic law for the electrons and the ions in the plane perpendicular to $\Bv$. This assumption is physically reasonable and leaves total freedom for what concerns the parallel polytropic laws for electrons and ions. The equilibrium condition for $\wgam=0$ reads
\begin{equation}\label{eq:EquilCond_fully-2}
\sqrt{f(x)^3} + \left[1-\wbetiprpz\FF(x)\right]\azi(x)f(x)
 - \sqrt{f(x)} - \azi(x) = 0\,,
\end{equation}
which can be interpreted as a cubic equation for $w(x)\equiv\sqrt{f(x)}$ ($f(x)$ cannot be negative since it is related to the pressure - see Eq.~(\ref{eq:MHD-FLR_functions})). In order to gain some insights from Eq.~(\ref{eq:EquilCond_fully-2}), we can solve it for $\azi(f)$, i.e.
\begin{equation}\label{eq:azi_f_implicit}
\azi(f) = \frac{\sqrt{f^3}-\sqrt{f\,}}{1-\left[1-\wbetiprpz\FF(x)\right]f}\,,
\end{equation}
together with the condition in Eq.~(\ref{eq:azi_condition}), which in our case, $h(x)=f(x)$, becomes
\begin{equation}\label{eq:azi_condition-2}
 \azi(f) \geq -\frac{\sqrt{f\,}}{2}\,.
\end{equation}
The implicit solution $\azi(f)$ is shown in Fig.~\ref{fig:figB0} for the case of $\FF=\HH=1$ and $|B_0|=1$, with different values of $\wbetiprpz$ and velocity shear strength, assuming a hyperbolic tangent profile of the type in Eq.~(\ref{eq:uiy_tanh}) with $L_u=3$. Three cases are shown: $\betiprpz=\beteprpz=0.1$ and $u_0=2/3$ (left panel), $\betiprpz=\beteprpz=1$ and $u_0=2/3$ (center), and $\betiprpz=\beteprpz=2$ and $u_0=1.5$ (right panel). The red continuous line represents the positivity condition border set by Eq.~(\ref{eq:azi_condition-2}), i.e. $\azi = -\sqrt{f\,}/2$, and only solutions above that curve are physical. The red dashed lines represent instead the maximum value of $\azi(x)$ for the chosen velocity profile.
\begin{figure}[!h]
  \begin{minipage}[!h]{0.3287\textwidth}
  \flushleft\includegraphics[width=1.0\textwidth]{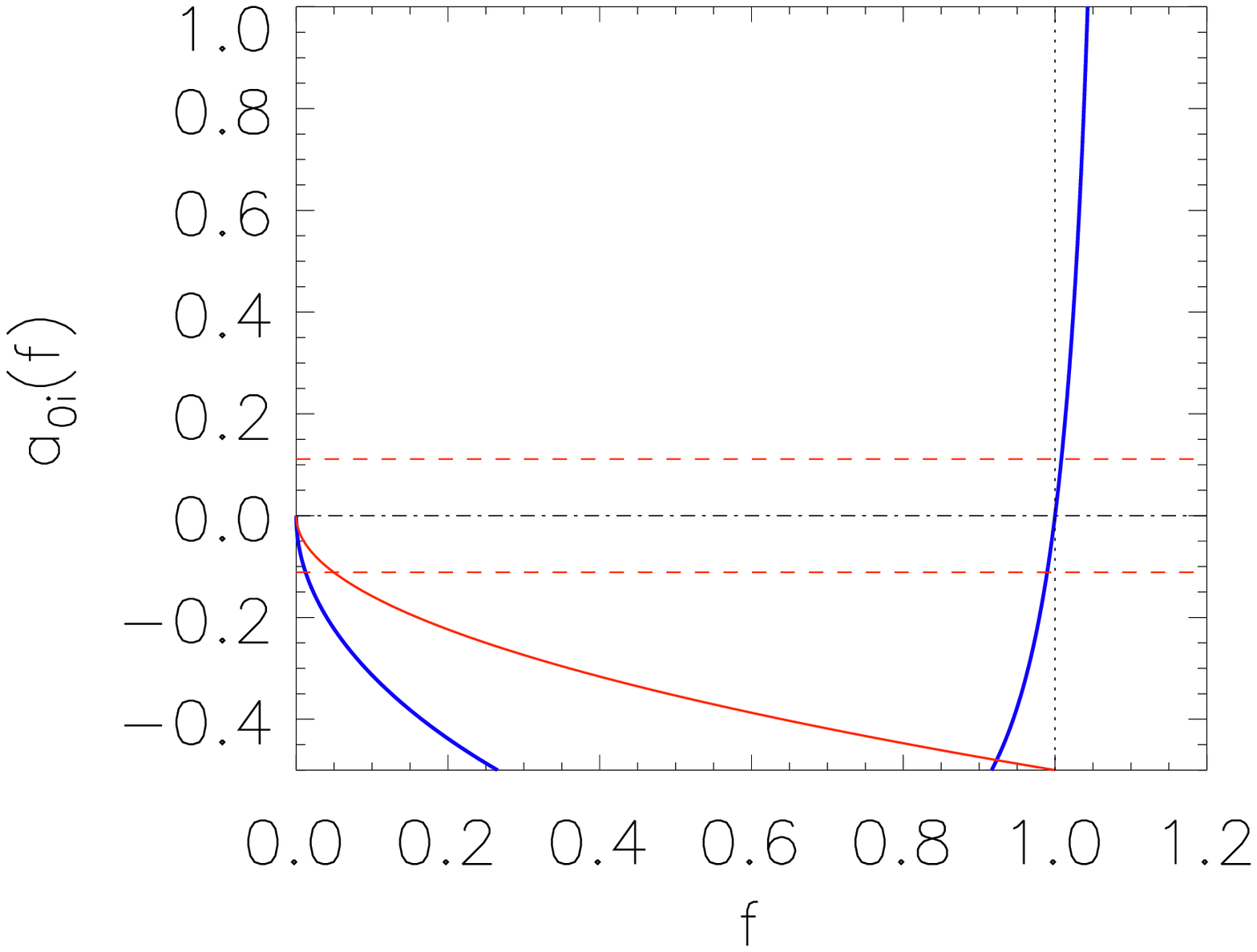}
  \end{minipage}
  \begin{minipage}[!h]{0.3287\textwidth}
  \flushleft\includegraphics[width=1.0\textwidth]{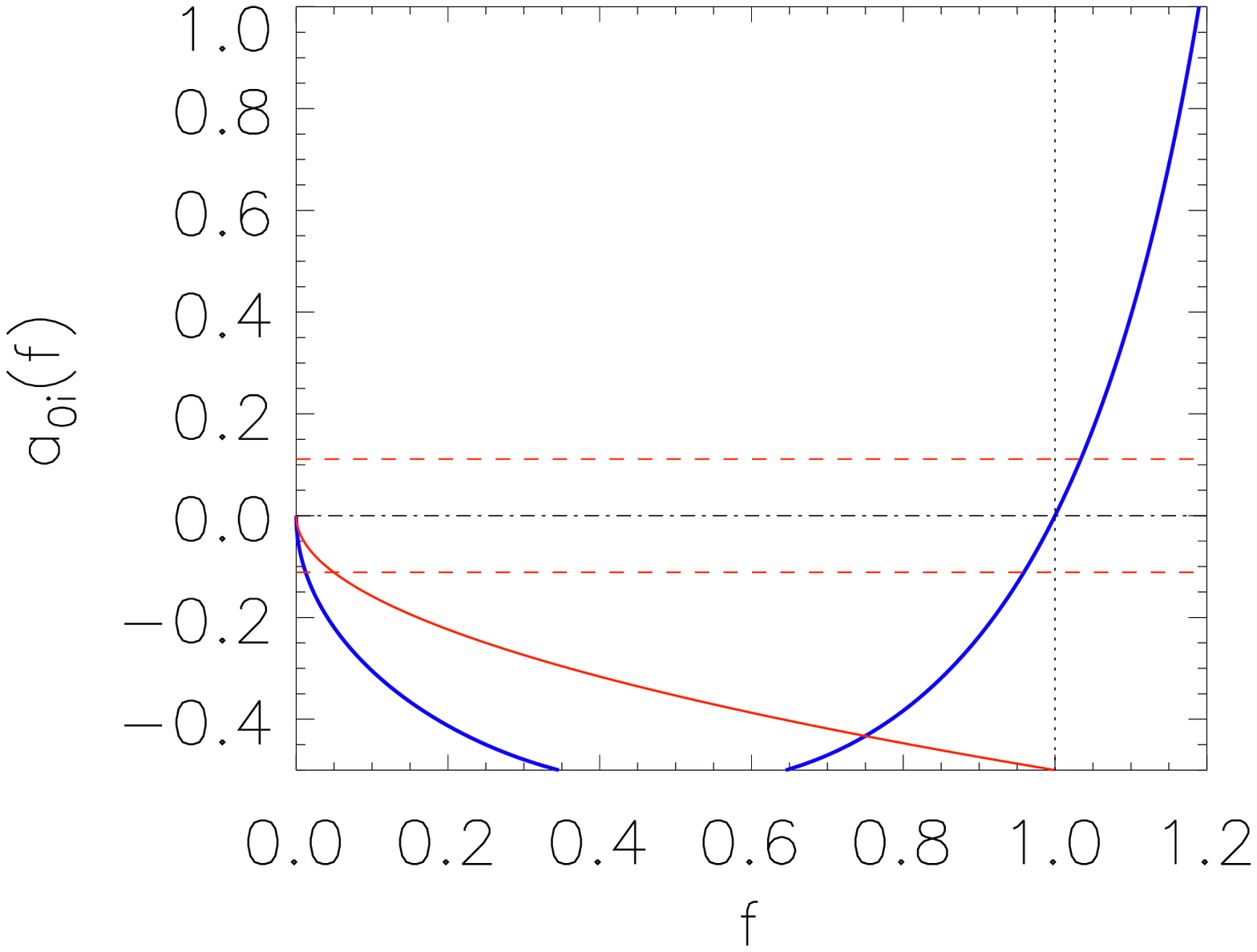}
  \end{minipage}
  \begin{minipage}[!h]{0.3287\textwidth}
  \flushleft\includegraphics[width=1.0\textwidth]{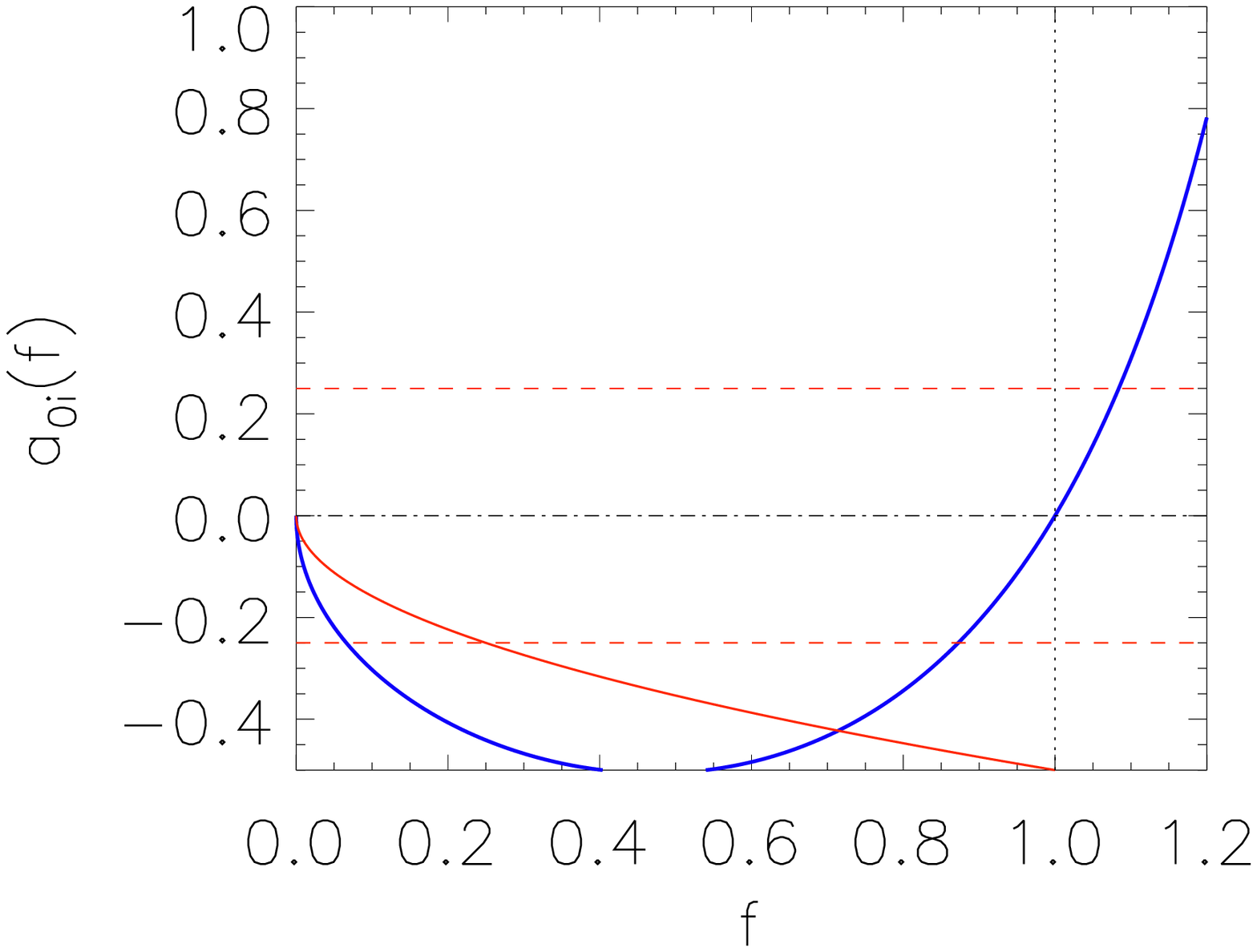}
  \end{minipage}
\caption{Plot of the implicit solution $\azi(f)$ in Eq.~(\ref{eq:azi_f_implicit}) for different plasma parameters: $\betiprpz=\beteprpz=0.1$ and $u_0=2/3$ (left), $\betiprpz=\beteprpz=1$ and $u_0=2/3$ (center), $\betiprpz=\beteprpz=2$ and $u_0=1.5$ (right). The other parameters are $\FF=\HH=1$, $|B_0|=1$, $L_u=3$ for all cases. Red continuous line represents the border above which the pressure is positive, Eq.~(\ref{eq:azi_condition-2}). Red dashed lines represent the bounds of $\azi$ values (for both $s_3=+1$ and $s_3=-1$), while the dotted and dash-dotted lines are the reference for $f=1$ and $\azi=0$, respectively}
\label{fig:figB0}
\end{figure} 
\\
From Fig.\ref{fig:figB0} several interesting features emerge: (i) the parameter $\wbetiprpz\FF$ (here $\FF=1$) essentially determines how ``fast'' the solution $f$ deviates from unity when $\azi$ deviates from zero, (ii) the strength of the velocity shear, i.e. the parameter $u_0/B_0L_u$, determines how far from gyrotropy the system is allowed to go, (iii) the presence of an asymmetry with respect to the sign of $\azi$ (due to the plot scale, this is more evident in the right panel, but it is true in general) and (iv) the existence of a second real solution for $\azi\leq0$ and not for $\azi>0$, which is however unphysical (i.e., below the positivity border - see below).
The above plots allow us to represent solutions in implicit form. Considering a hyperbolic tangent velocity shear profile as in Eq.~(\ref{eq:uiy_tanh}) and $\HH=1$, the function $\azi(x)$ reads
\begin{equation}\label{eq:azi_x_tanh}
\azi(x) = \frac{s_3}{2}\frac{u_0}{B_0L_u}\cosh^{-2}\left(x/L_u\right)\,,
\end{equation}
where we took $x_0=0$ for simplicity. We consider the case shown in the central panel of Fig.~\ref{fig:figB0} for $\azi(f)$. This plot of $\azi$ versus $f$, zoomed  in the region around  $\azi=0$ and $f=1$, is reproduced in the right panel of Fig.~\ref{fig:figB1}. In the left panel of Fig.~\ref{fig:figB1}, we show the plot of $\azi(x)$ in Eq.~(\ref{eq:azi_x_tanh}), for both signs of $s_3$. The aim is to visualize how $f(x)$ should look by drawing  $\azi(x)$ and $\azi(f)$ next to  each other, with the values of $\azi$ on the $y$-axis of both plots (and with the same scale), so one can go back and forth from one plot to the other in three steps: go along the $x$-axis of left panel of Fig.~\ref{fig:figB1} ($x$ coordinate), (ii) look at the value that $\azi(x)$ takes in the left panel and then move to the same level of right panel (in both panels, the $y$-axis has the same scale and the values of $\azi$ are represented on it) and (iii) after reaching  the curve $\azi(f)$ on the right panel in the point that corresponds to the value of $\azi$ found at point (ii), go down to the $x$-axis of the right panel in order to find  the value of $f$ that corresponds to the value of $x$ at point (i).
\begin{figure}[!h]
  \begin{minipage}[!h]{0.49\textwidth}
  \flushleft\includegraphics[width=1.0\textwidth]{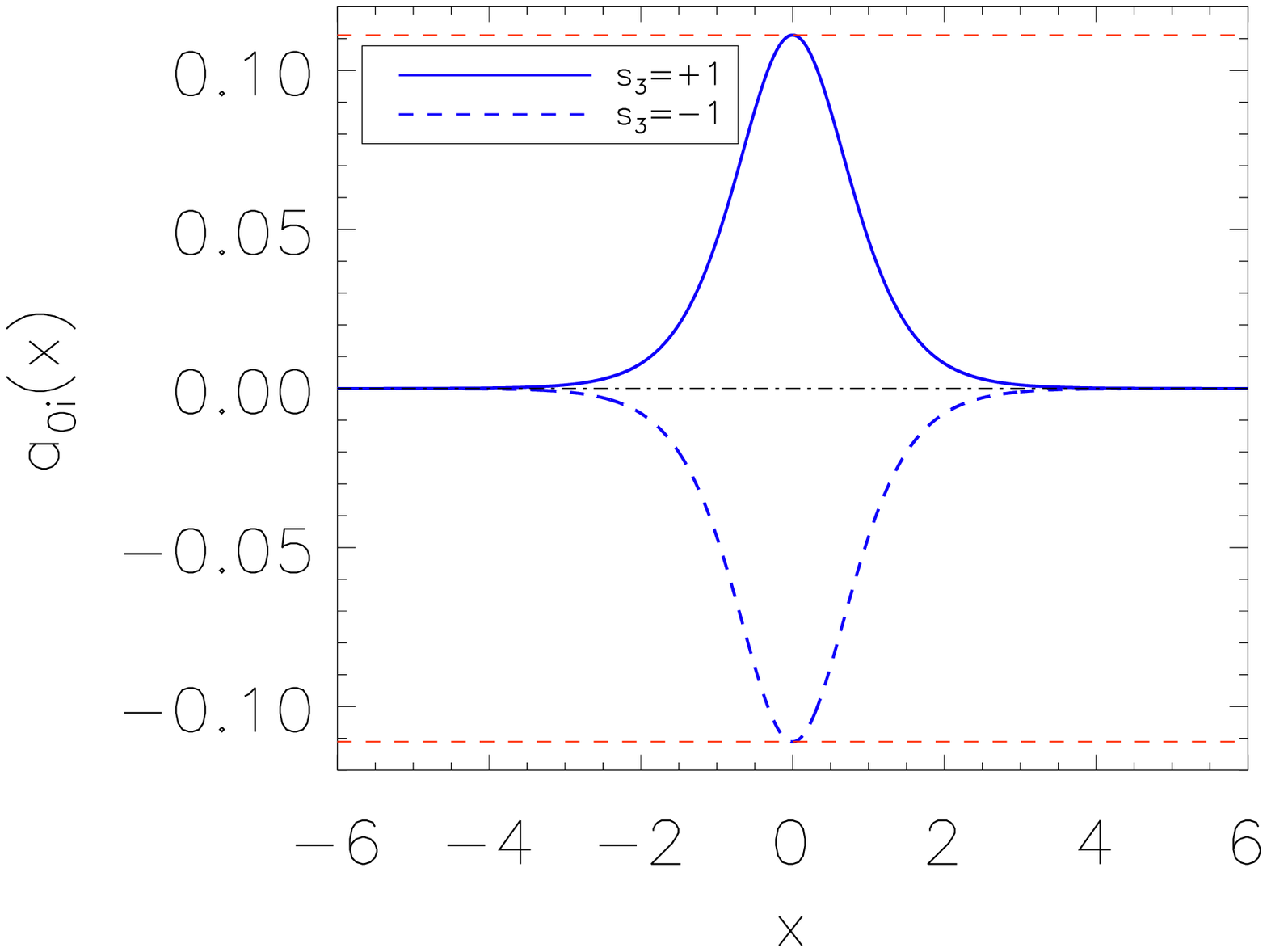}
  \end{minipage}
  \begin{minipage}[!h]{0.49\textwidth}
  \flushleft\includegraphics[width=1.0\textwidth]{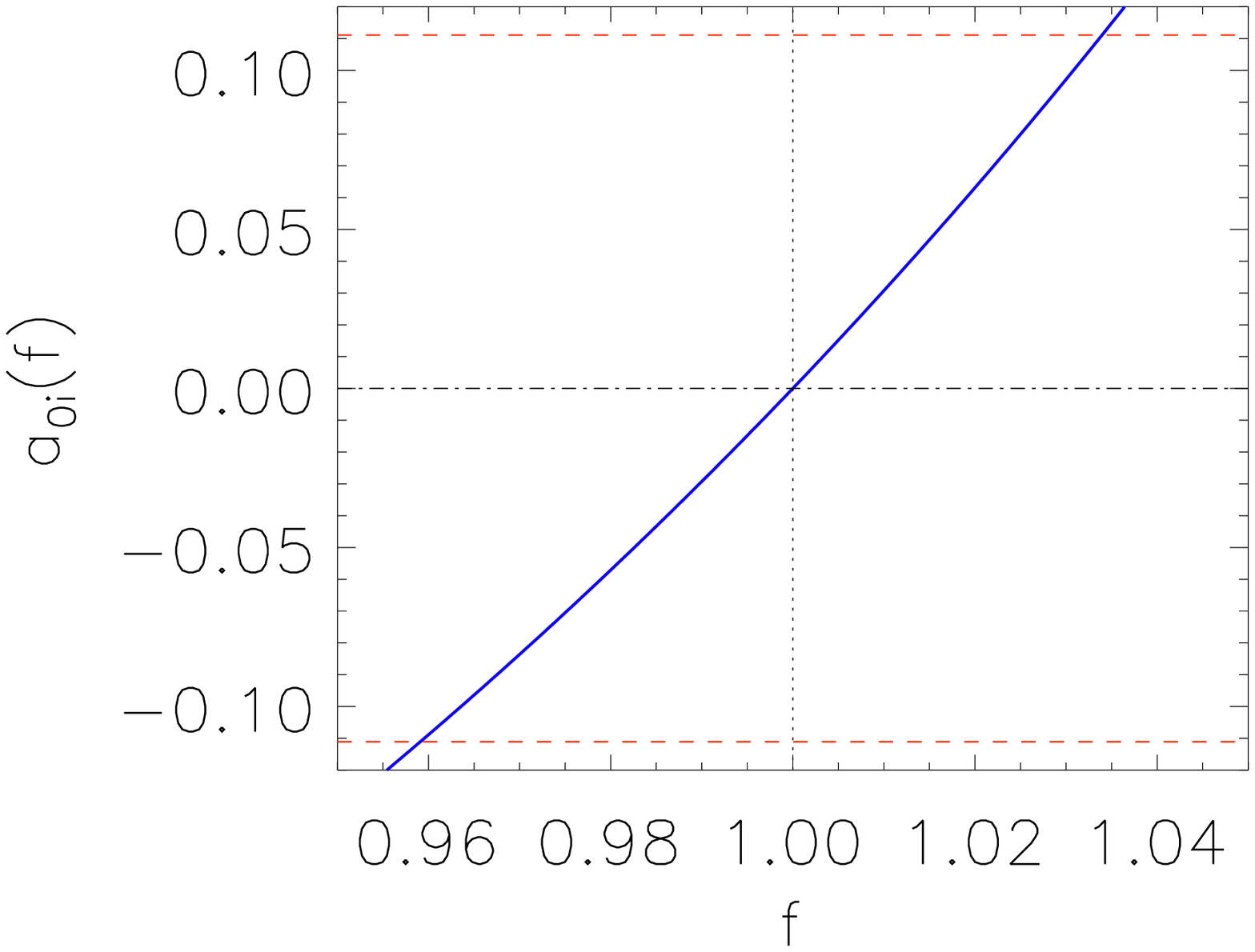}
  \end{minipage}
\caption{Left: plot of $\azi(x)$ in Eq.~(\ref{eq:azi_x_tanh}) for both sign of $s_3$ (see {\bf insert}). Right: zoom around $f=1$ and $\azi=0$ of the center plot in Fig.~\ref{fig:figB0}.}
\label{fig:figB1}
\end{figure} 
\\
Imagine going along the whole $x$-axis, from $-\infty$ to $+\infty$. For sufficiently large $|x|$, we have $\azi=0$, so the solution is asymptotically $f=1$ (e.g. for $|x|\gtrsim4$ in left panel of Fig.~\ref{fig:figB1}). Then, still 
referring to  Fig.~\ref{fig:figB1}, as we approach $x=0$ from negative values $\azi$ starts to deviate from zero (left panel) and thus also $f(x)$  starts to deviate from unity (right panel), becoming bigger or smaller if $\azi$ becomes positive or negative, respectively. In passing through $x=0$ (left panel), we pass through the global maximum (minimum) of the positively (negatively) valued $\azi$, corresponding to the maximum deviation of $f(x)$ from unity in the right panel (i.e., in the point where the curve $\azi(f)$ intercepts the horizontal red dashed line, above $f=1$ or below it accordingly with the sign of $\azi$). Then, leaving $x=0$ behind and proceeding to increasing positive $x$-values in the left panel of Fig.~\ref{fig:figB1}, $\azi$ starts to decrease (increase) and so does $f(x)$ in the right panel, until it comes back to unity for sufficiently high $x$-values (note that for $x>0$ we are going on the curve $\azi(f)$ all the way back compared to how it had been covered for $x<0$, the point $x=0$ being the turning point).\\
Explicit numerical solutions $f(x)$ of Eq.~(\ref{eq:EquilCond_fully-2}) for $\azi(x)$ given in Eq.~(\ref{eq:azi_x_tanh}), are plotted in Fig.~\ref{fig:figB3} for the three cases in Fig.~\ref{fig:figB0}. The corresponding profiles $\Pixx(x)$ ($s_3=+1$: bottom blue solid line, $s_3=-1$: top blue dashed line) and $\Piyy(x)$ ($s_3=+1$: top red solid line, $s_3=-1$: bottom red dashed line) are also given (from Eqs.~(\ref{eq:FLR_equil_Pi_xx}) and (\ref{eq:FLR_equil_Pi_yy}) - see below). The asymmetry with respect to the sign of $\Omegav_\uv\cdot\Bv$ is more evident on the right panel due to the choice of the parameters, but it is present in all cases.
\begin{figure}[!h]
  \begin{minipage}[!h]{0.3287\textwidth}
  \flushleft\includegraphics[width=1.0\textwidth]{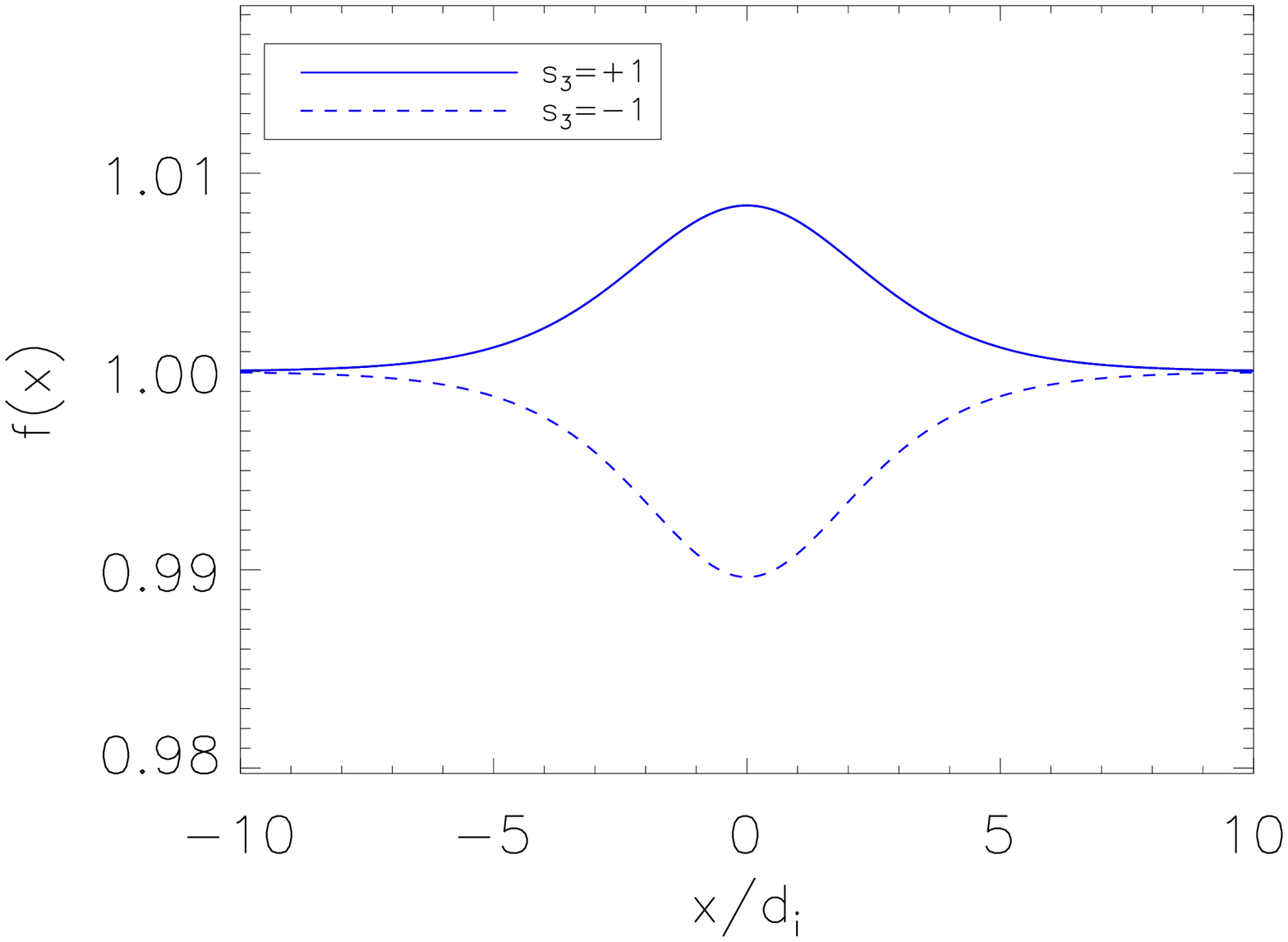}
  \end{minipage}
  \begin{minipage}[!h]{0.3287\textwidth}
  \flushleft\includegraphics[width=1.0\textwidth]{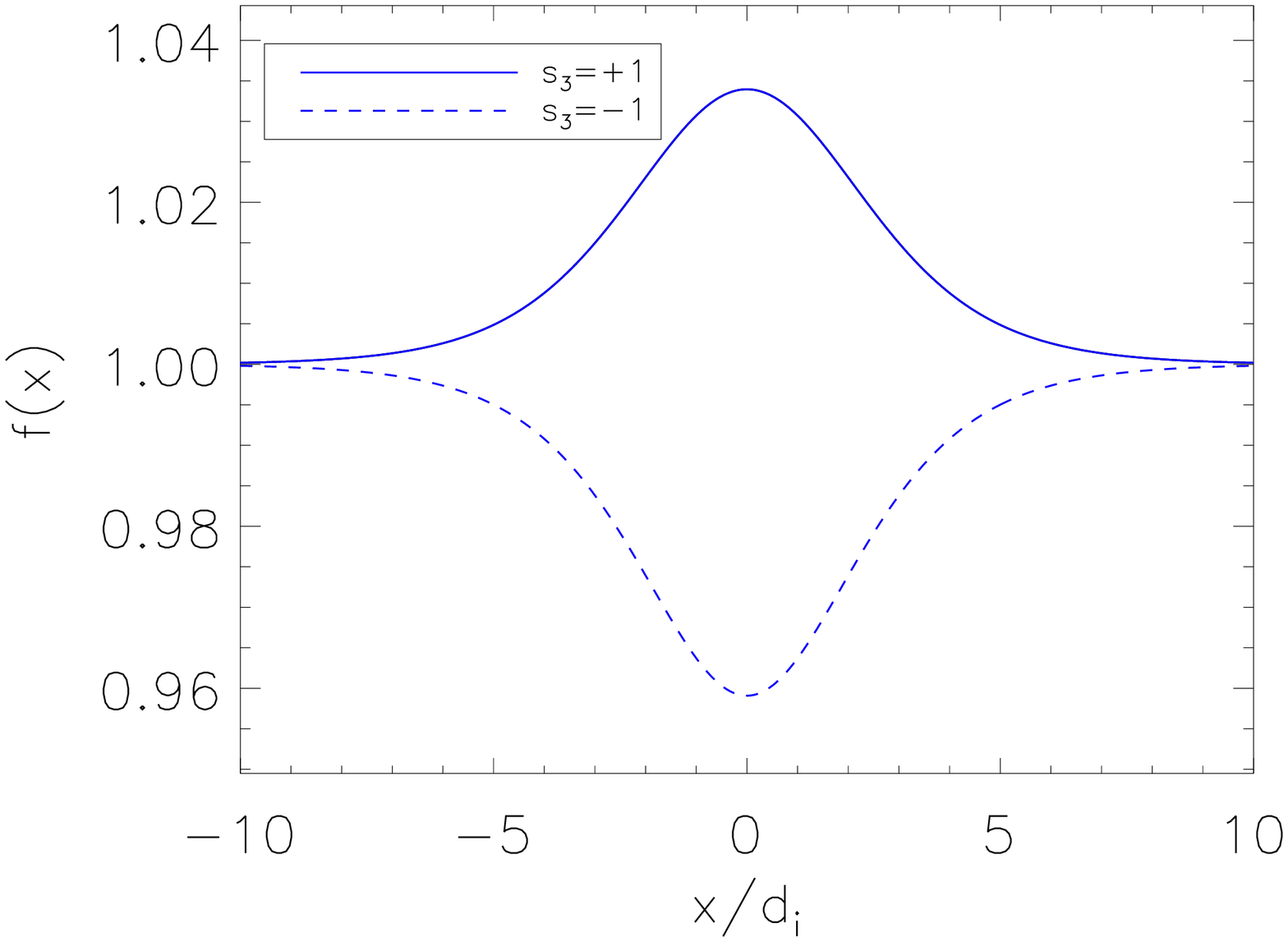}
  \end{minipage}
  \begin{minipage}[!h]{0.3287\textwidth}
  \flushleft\includegraphics[width=1.0\textwidth]{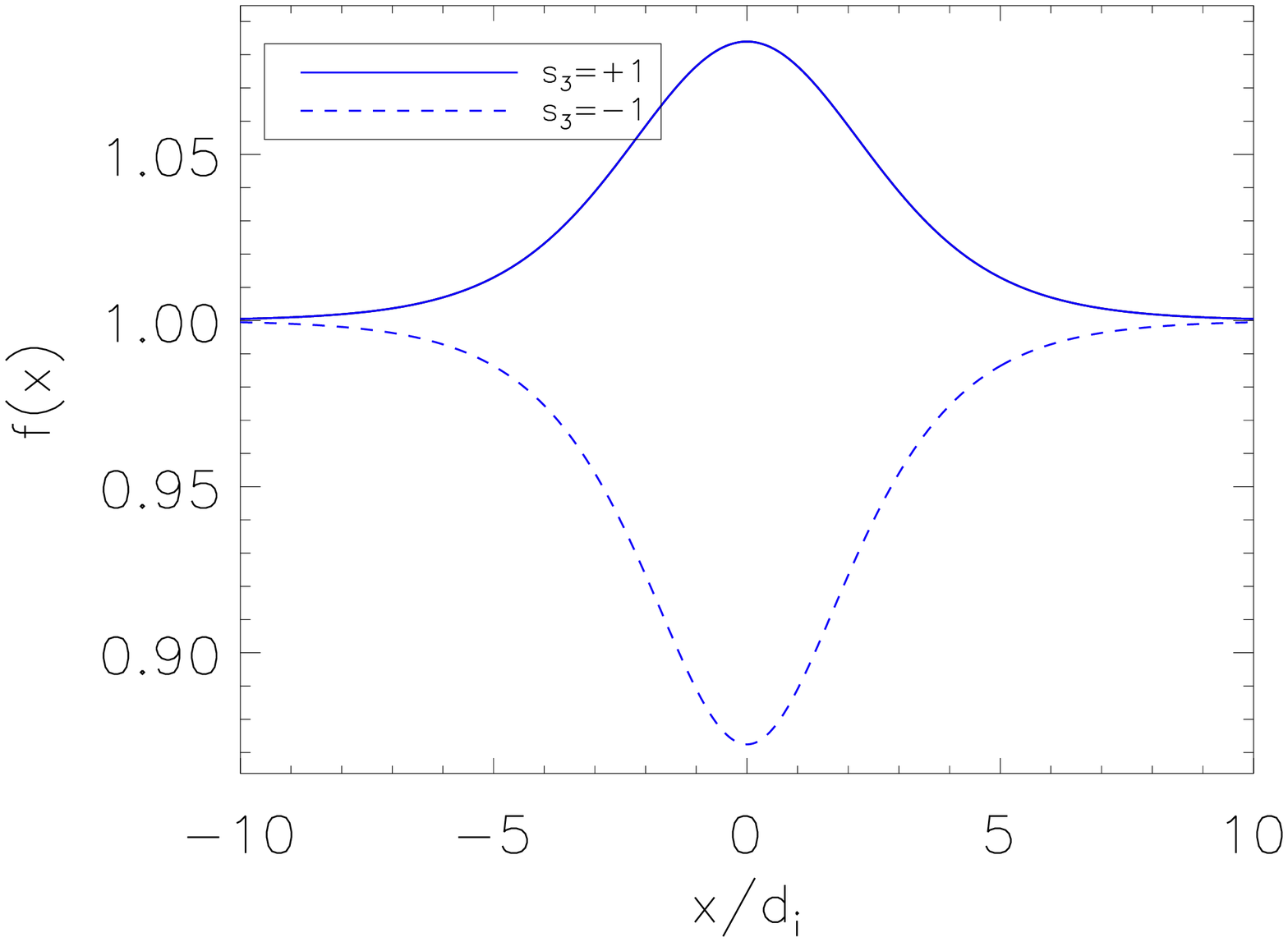}
  \end{minipage}\\
  \begin{minipage}[!h]{0.3287\textwidth}
  \flushleft\includegraphics[width=1.0\textwidth]{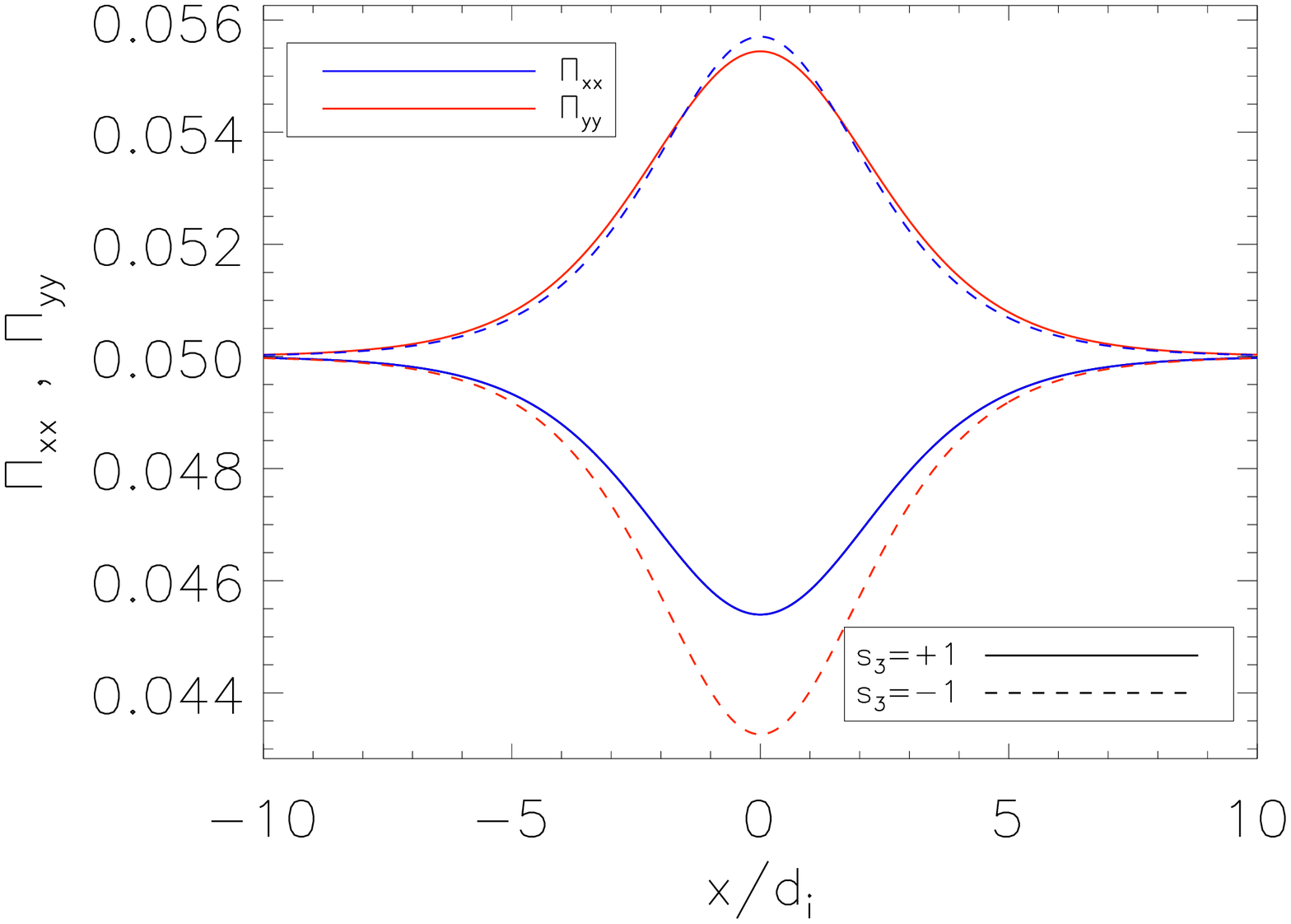}
  \end{minipage}
  \begin{minipage}[!h]{0.3287\textwidth}
  \flushleft\includegraphics[width=1.0\textwidth]{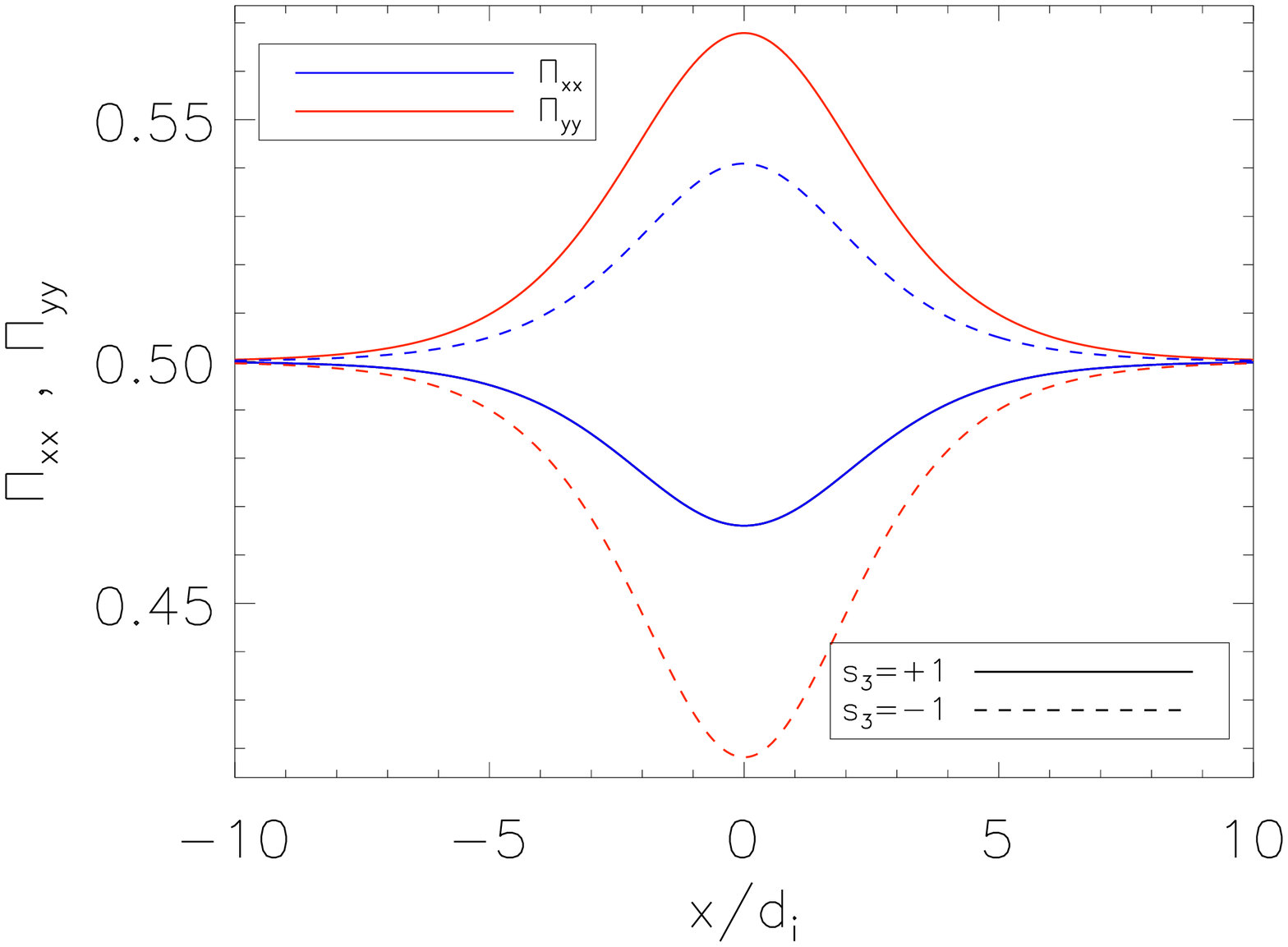}
  \end{minipage}
  \begin{minipage}[!h]{0.3287\textwidth}
  \flushleft\includegraphics[width=1.0\textwidth]{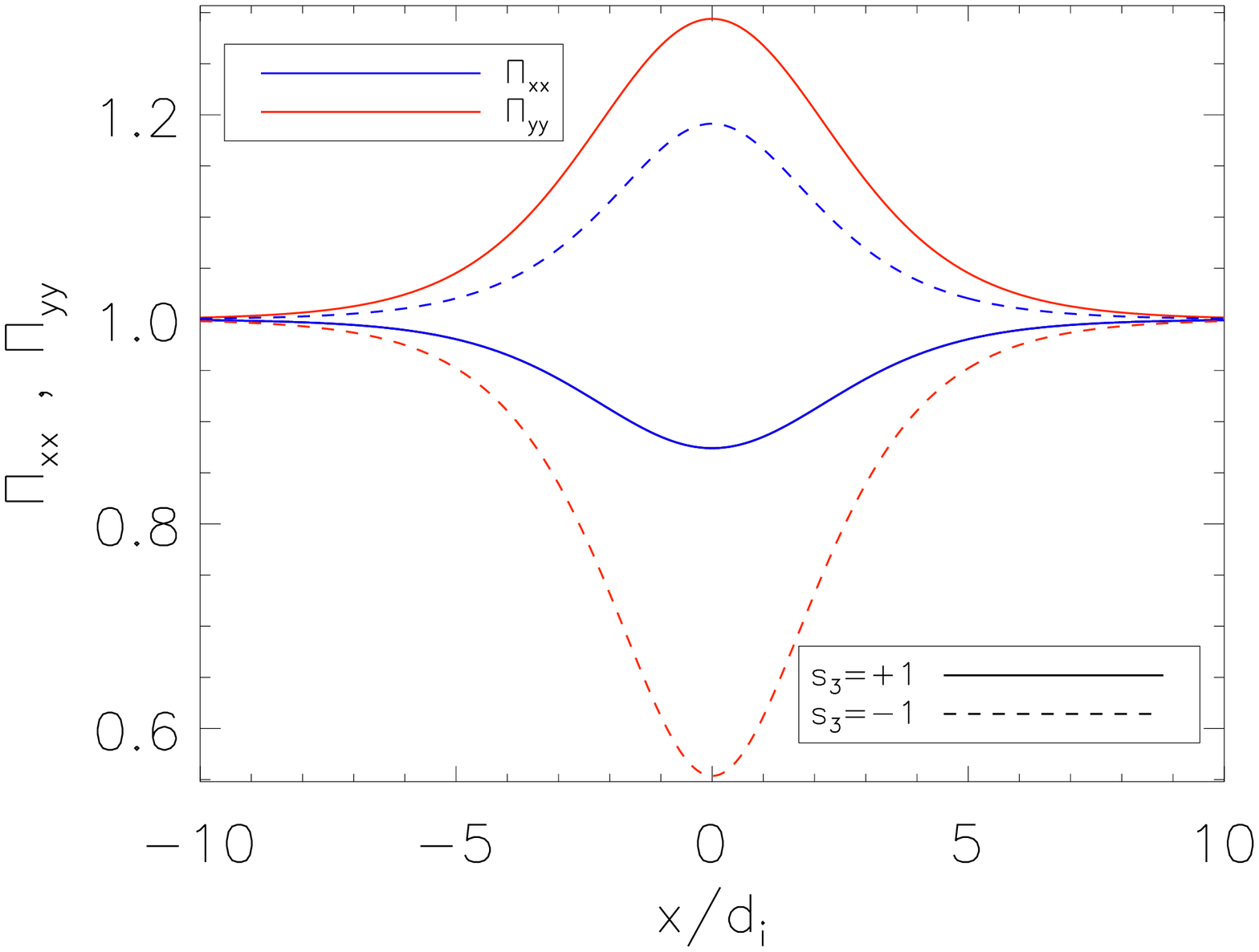}
  \end{minipage}
\caption{Top row: plot of the explicit numerical solution $f(x)$ of Eq.~(\ref{eq:EquilCond_fully-2}). Bottom row: corresponding $\Pixx$ and $\Piyy$ profiles (see in the text). The three cases above correspond to the three cases in Fig.~\ref{fig:figB0}: $\betiprpz=\beteprpz=0.1$ and $u_0=2/3$ (left), $\betiprpz=\beteprpz=1$ and $u_0=2/3$ (center), $\betiprpz=\beteprpz=2$ and $u_0=1.5$ (right).}
\label{fig:figB3}
\end{figure} 
\\
In order to trace back the origin of the double solution for $\azi\leq0$ and to show  that one of the two solutions is  unphysical because of the positivity condition in Eq.(\ref{eq:azi_condition-2}), we consider the equilibrium condition, Eq.~(\ref{eq:EquilCond_fully-2}), and look for solutions which deviate from gyrotropy weakly. For this purpose, we  treat $\wbetiprpz$ as a small parameter (here, we consider the case $\FF=1$, but the bound $0\leq\FF\leq1$ always holds). For $\wbetiprpz=0$, Eq.~(\ref{eq:EquilCond_fully-2}) is exactly solvable:
\begin{equation}\label{eq:EquilCond_fully_f0}
\left(f(x) - 1\right)\left(\sqrt{f(x)} + \azi(x)\right) = 0\,,
\end{equation}
and  admits  two real roots
\begin{equation}\label{eq:Sol_fully_f0}
\left\{\begin{array}{ccc}
f_0(x) = 1 & &\forall\ x\\
\tilde{f}_0(x) = \azi^2(x)& & \forall\ x\in\left\{x | \azi(x)\leq0\right\}
\end{array}\right.
\end{equation}
where $f_0=1$ means gyrotropy, while $\tilde{f}_0$ represents the $\wbetiprpz\to0$ limit of the second solution in Fig.~\ref{fig:figB0} for $\azi\leq0$. However, $\tilde{f}_0$ is  below the positivity condition $\ai\geq-1/2$ (for $f\to\tilde{f}_0$ we obtain  $\ai\to-1$). Keeping  the first order in $\wbetiprpz\FF\ll1$ for the physical solution near unity leads to 
\begin{equation}\label{eq:Sol_f0-f1}
f(x)\simeq 1 + \wbetiprpz\FF(x)\frac{\azi(x)}{1+\azi(x)}\,,
\end{equation}
which corresponds to the first-order Taylor expansion for small $\CC_0\azi/(1+\azi)$ of the solution in Eq.~(\ref{eq:FLR_equil_0}) (see below). This means that, at least in the limit of small-$\wbetiprpz$, the two solutions are close to each other. Note that the small-$\wbetiprpz$ limit does not necessarily mean small-$\betiprpz$, but it can be reached also in the large perpendicular temperature ratio, $\tauprp=\Teperpz/\Tiperpz\gg1$.
Finally, physical solutions of Eq.~(\ref{eq:EquilCond_fully-2}) only exist within restricted domains of  ($\wbetiprpz$, $\azi$). In particular, this turns out to be the case only when $\azi$ is negative, i.e. for $\Omegav_{\uv}\cdot\Bv<0$. An example of this asymmetric behavior is given in Fig.~\ref{fig:figB2}, where  we show that  solutions for $\azi<0$  may disappear  because of the positivity constraint depending on the value of $\wbetiprpz$. For $\azi>0$, i.e. for $\Omegav_{\uv}\cdot\Bv>0$, a solution is instead always present, regardless of the value of $\wbetiprpz$.
\begin{figure}[!h]
  \begin{minipage}[!h]{0.49\textwidth}
  \flushleft\includegraphics[width=1.0\textwidth]{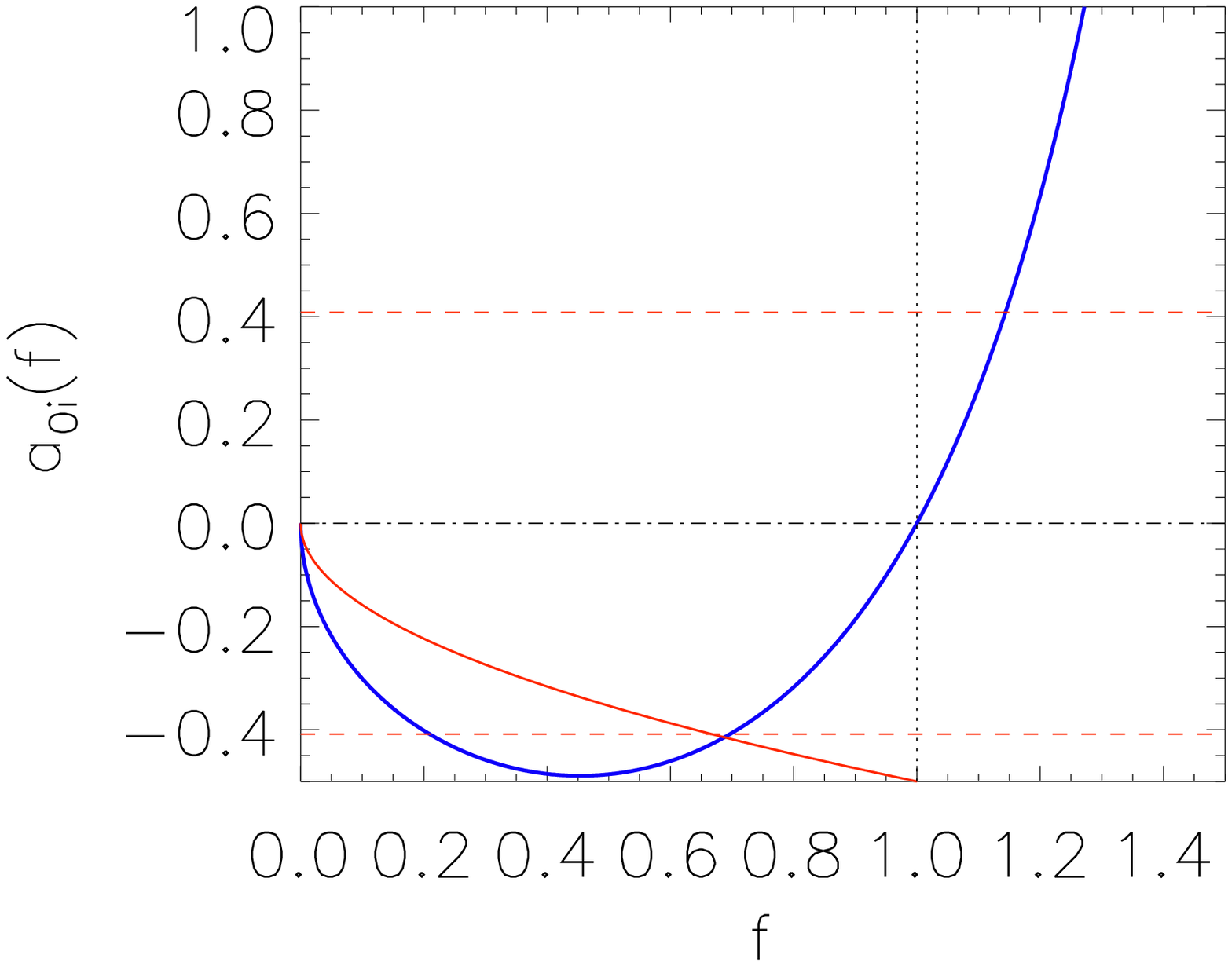}
  \end{minipage}
  \begin{minipage}[!h]{0.49\textwidth}
  \flushleft\includegraphics[width=1.0\textwidth]{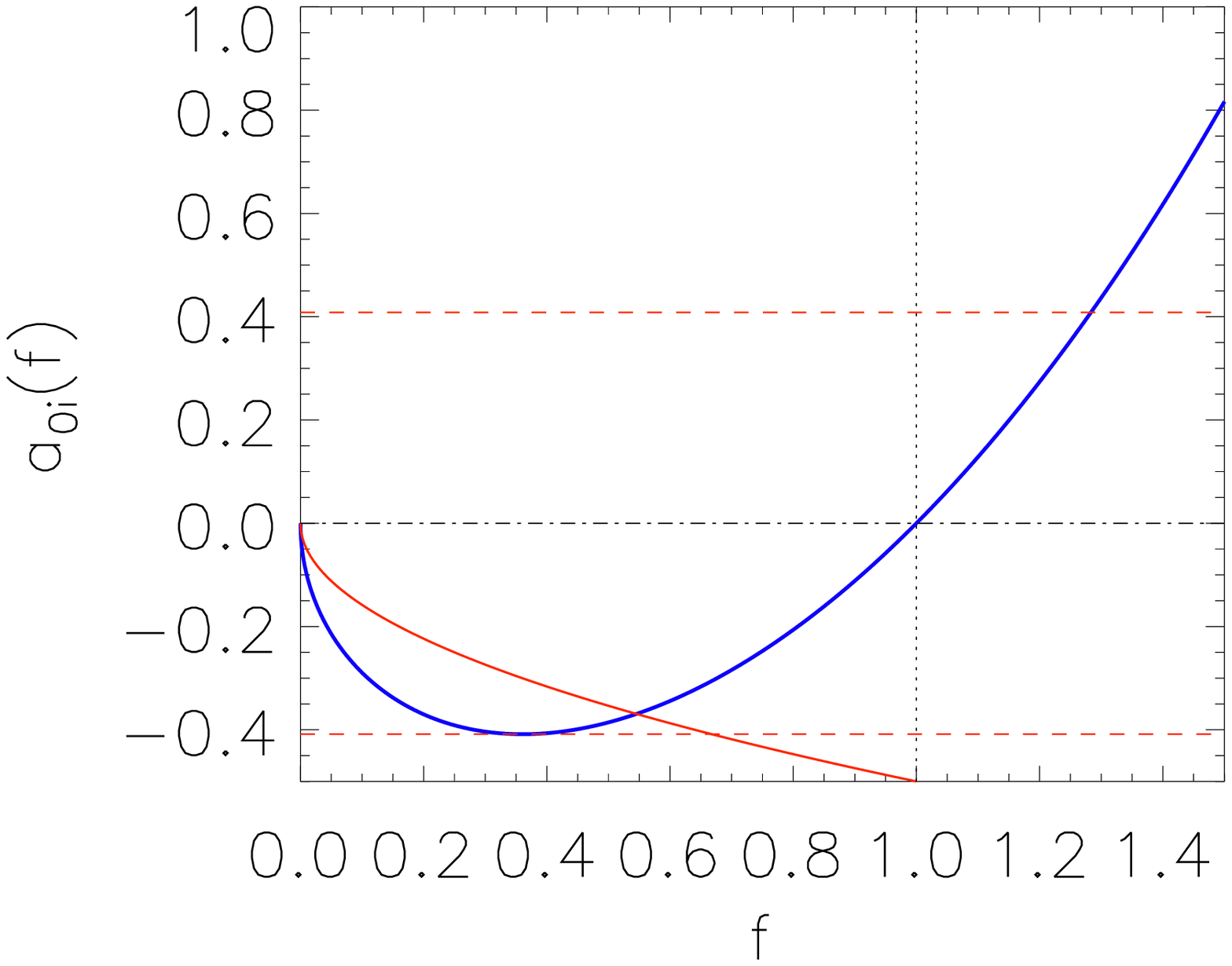}
  \end{minipage}
\caption{Plot of the implicit solution $\azi(f)$ for the case $\betiprpz=\beteprpz=5$ (left) and for the case $\betiprpz=10$, $\beteprpz=1$ (right), corresponding to $\wbetiprpz=5/11$ and $5/6$, respectively. The other parameters are $\FF=\HH=1$, $u_0=2.45$, $|B_0|=1$, $L_u=3$ for both cases.}
\label{fig:figB2}
\end{figure} 
\\

\subsection{Approximate analytical solution of the equilibrium condition}\label{subsec:FLRequilibria_approx}

In the limit of small corrections to the MHD equilibrium, we can give analytical solutions for $f(x)$ with $\wgam\neq0$. In this limit, the ion cyclotron frequency $\Omega_{\rm i}$ in the $\ai$ function can be computed with the MHD profile of the magnetic field, $|\Bv(x)|\simeq B_0\sqrt{\HH(x)}$, i.e.
\begin{equation}\label{eq:a_meanB}
 \ai(x) \simeq \azi(x) = \frac{s_3}{2B_0\sqrt{\HH(x)}}\duiydx\,,
\end{equation}
which is an approximation that one should check a posteriori (see Appendix \ref{app:FLRequilibria_altern} for a simple case in which this case is exact), and leads to the following equilibrium condition:
\begin{equation}\label{eq:FLR_equil_cond_Bmean}
 \left[1 - \wbetiprpz\FF(x)\frac{\azi(x)}{1+\azi(x)} +
 \wbeteprpz\left(\FF(x)\right)^{1+\wgam}\left(f^{\wgam}(x)-1\right)\right]f(x) = 1\,.
\end{equation}
Treating $\wgam$ as a small parameter, we solve the above equilibrium problem iteratively. The solution of Eq.~(\ref{eq:FLR_equil_cond_Bmean}) for $\wgam=0$ is straightforward, i.e.
\begin{equation}\label{eq:FLR_equil_0}
 f_0(x) = \left[1-\CC_0(x)\frac{\azi(x)}{1+\azi(x)}\right]^{-1}\,,
\end{equation}
where $\CC_0(x)=\wbetiprpz\FF(x)$. Noting that Eq.~(\ref{eq:FLR_equil_cond_Bmean}) is equivalent to the equilibrium condition Eq.~(23) in Ref.~\cite{CerriPOP2013}, with the substitution
\[
 \widetilde{u}'(x)\,\,\,\to\,\,\, \frac{\azi(x)}{1+\azi(x)}\,,
\]
we can derive the iterative solution (cf. Eqs.(24)-(26) in Ref.~\cite{CerriPOP2013}), i.e. 
\begin{equation}\label{eq:FLR_equil_conv}
 f(x) = \left[1-\CC(x)\frac{\azi(x)}{1+\azi(x)}\right]^{-1}\,,
\end{equation}
with 
\begin{equation}\label{eq:FLR_coeff_conv}
 \CC(x) = \left[1+\wgam\wbeteprpz\left(\FF(x)\right)^{1+\wgam}\right]^{-1}\CC_0(x)\,,
\end{equation}
In the above, we have assumed that $|\wgam\wbeteprpz(\FF(x))^{1+\wgam}|<1$ $\forall x$ for the convergence of the resulting series, which can be shown to be always the case (see Ref.~\cite{CerriPOP2013}). Moreover, the solution passes through a Taylor expansion in which we consistently assume that the correction to the MHD profile is small, i.e. $|\CC\azi/(1+\azi)|\ll1$. In this regard, the relation $|\azi/(1+\azi)|\leq1$ $\forall\ \azi\geq-1/2$ holds, $\wbetiprpz<1$ by definition and we can always choose $\piperpz$ such that $\FF(x)<1$ $\forall x$. Thus, the condition $|\CC\azi/(1+\azi)|\ll1$ is valid for most of the parameter range commonly adopted. Finally, the solution in Eq.~(\ref{eq:FLR_equil_conv}) reduces to the first-order FLR solution given in Ref.~\cite{CerriPOP2013} if we retain only the first order in $\azi$ inside the square brackets.

\subsubsection{Explicit equilibrium profiles}

We give also the explicit equilibrium profiles of the physical quantity of interest, for the sake of clarity:
\begin{equation}\label{eq:FLR_equil_Pi_xx}
 \Piixx\ =\ \piperpz\left(1-\frac{\ai(x)}{1+\ai(x)}\right)\FF(x)f(x)\,,
\end{equation}
\begin{equation}\label{eq:FLR_equil_Pi_yy}
 \Piiyy\ =\ \piperpz\left(1+\frac{\ai(x)}{1+\ai(x)}\right)\FF(x)f(x)\,,
\end{equation}
\begin{equation}\label{eq:FLR_equil_Pi_zz}
 \Piizz\ \equiv\ \pipara\ =\ \piparaz\left(\FF(x)f(x)\right)^{\gipara/\giperp}\,,
\end{equation}
\begin{equation}\label{eq:FLR_equil_den}
 n(x)\ =\ n_0\left(\FF(x)f(x)\right)^{1/\giperp}\,,
\end{equation}
\begin{equation}\label{eq:FLR_equil_peperp}
 \Piexx\ = \Pieyy\ \equiv\ \peperp =\ \peperpz\left(\FF(x)f(x)\right)^{1+\wgam}\,, 
\end{equation}
\begin{equation}\label{eq:FLR_equil_pepara}
 \Piezz\ \equiv\ \pepara\ =\ \peparaz\left(\FF(x)f(x)\right)^{\gepara/\giperp}\,,
\end{equation}
\begin{equation}\label{eq:FLR_equil_Bz}
 B_z(x)\ =\ B_0\sqrt{\HH(x)f(x)}\,.
\end{equation}
If the parallel and perpendicular temperatures are of interest~\cite{PassotPOP2006}, instead of the corresponding  pressures,  from Eqs.~(\ref{eq:FLR_equil_Pi_xx})-(\ref{eq:FLR_equil_den})  we have
\begin{equation}\label{eq:FLR_equil_Tiperp}
 \Tiperp\ \equiv\ \frac{1}{n}\frac{\Piiyy+\Piiyy}{2}\ =\ \Tiperpz\left(\FF(x)f(x)\right)^\frac{\giperp-1}{\giperp}\,,
\end{equation}
\begin{equation}\label{eq:FLR_equil_Tipara}
 \Tipara\ \equiv\ \frac{\pipara}{n}\ =\ \Tiparaz\left(\FF(x)f(x)\right)^\frac{\gipara-1}{\giperp}\,, 
\end{equation}
\begin{equation}\label{eq:FLR_equil_Tiperp}
 \Teperp\ \equiv\ \frac{\peperp}{n}\ =\ \Tiperpz\left(\FF(x)f(x)\right)^\frac{\geperp-1}{\giperp}\,,
\end{equation}
\begin{equation}\label{eq:FLR_equil_Tipara}
 \Tepara\ \equiv\ \frac{\pepara}{n}\ =\ \Tiparaz\left(\FF(x)f(x)\right)^\frac{\gepara-1}{\giperp}\,.
\end{equation}
We define the MHD profiles with the  superscript $(0)$, e.g. $\Tiperp^{(0)}\equiv\Tiperpz\FF(x)^{(\giperp-1)/\giperp}$, and the overlined quantities  as the ratio between the actual and the MHD  profiles, e.g. $\bar{T}_{{\rm i},\perp}\equiv\Tiperp/\Tiperp^{(0)}$. Note that taking the perpendicular and parallel polytropic indices to be $\gamma_\perp=2$ and $\gamma_\|=1$ for both species, one obtains 
\begin{equation}\label{eq:FLR_equil_Tpara-double-adiab}
 \bar{T}_\|\ =\ 1\,,
\end{equation}
\begin{equation}\label{eq:FLR_equil_Tipara}
 \bar{T}_\perp\ =\ \bar{n}\ =\ \bar{B}\,, 
\end{equation}
which mean $\bar{p}_\|\bar{B}^2/\bar{n}^3 = {\rm const.}$ and $\bar{p}_\perp/\bar{n}\bar{B} = {\rm const.}$. Thus, if the initial MHD profiles are such that $p_\|^{(0)}(B^{(0)})^2/(n^{(0)})^3 = {\rm const.}$ and $p_\perp^{(0)}/n^{(0)}B^{(0)} = {\rm const.}$ (as, e.g., for the case $\FF=\HH=1$), then also $p_\|B^2/n^3 = {\rm const.}$ and $p_\perp/nB = {\rm const.}$, as expected from the double adiabatic laws~\cite{CGL1956}.\\
Finally, in order to  visualize the asymmetry due to the sign of $\Omegav_\uv\cdot\Bv$, we  plot the equilibrium profiles for a given case. In Fig.~\ref{fig:fig1} we show the velocity shear $\uiy(x)$ and the function $\CC(x)\azi(x)/[1+\azi(x)]$ (left panel) and the corresponding equilibrium profiles for $\Piixx$ and $\Piiyy$ (right panel).
\begin{figure}[!h]
  \begin{minipage}[!h]{0.49\textwidth}
  \flushleft\includegraphics[width=1.0\textwidth]{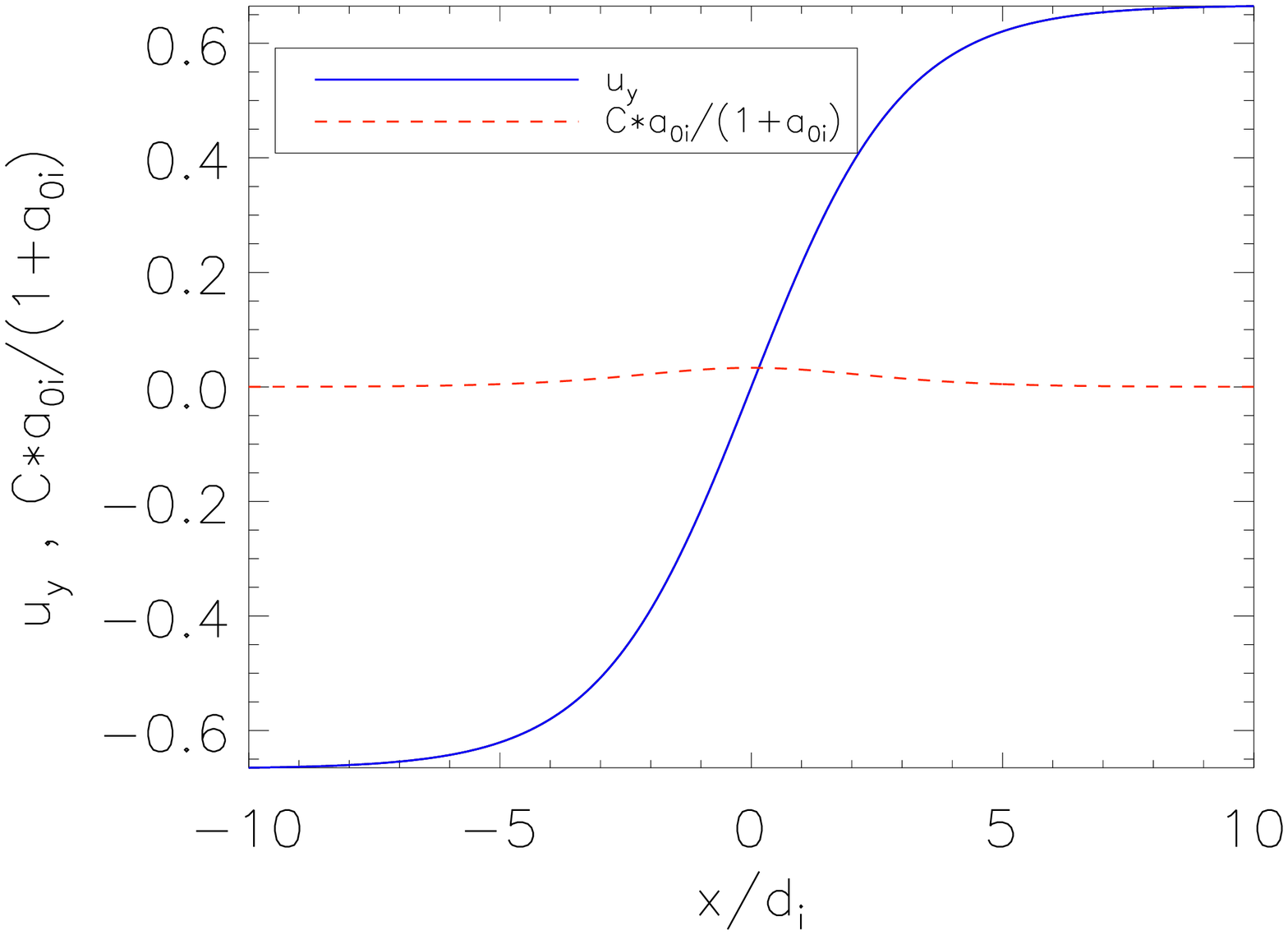}
  \end{minipage}
  \begin{minipage}[!h]{0.49\textwidth}
  \flushleft\includegraphics[width=1.0\textwidth]{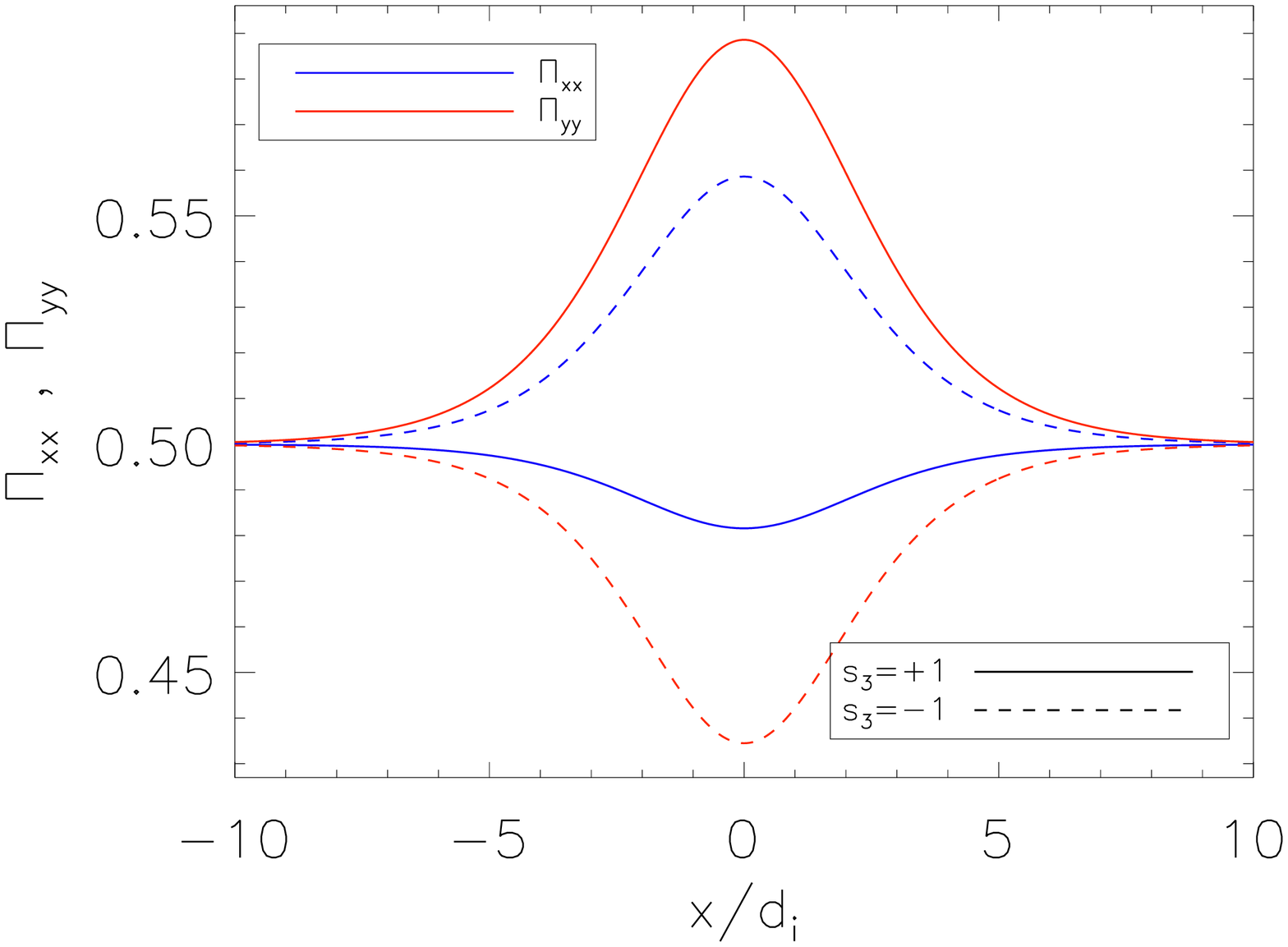}
  \end{minipage}
\caption{Left: velocity profile $\uiy(x)$ (blue solid line) and the function $\CC(x)\ai(x)/[1+\ai(x)]$ (red dashed line). Right: plot of the approximated equilibrium profiles $\Piixx(x)$ ($s_3=+1$: bottom blue solid line, $s_3=-1$: top blue dashed line) and $\Piiyy(x)$ ($s_3=+1$: top red solid line, $s_3=-1$: bottom red dashed line). Here, the parameters are: $\FF=\GG=\HH=1$, $u_0=2/3$, $L_u=3$, $B_0=\pm1$ ($s_3=\pm1$), $\betiprpz=\beteprpz=1$ and $\wgam=0$.}
\label{fig:fig1}
\end{figure} 
The parameters used for the profiles in Fig.~\ref{fig:fig1} are $u_0=2/3$, $L_u=3$, $B_0=\pm1$ ($s_3=\pm1$), $\betiprpz=\beteprpz=1$ and $\wgam=0$. For the sake of clarity we have chosen the simplest MHD case of $\FF=\GG=\HH=1$. By an inspection of the plot, some considerations emerge. First, while the condition $|\CC(x)\azi(x)/[1+\azi(x)]|\ll1$ holds, as assumed in the derivation on the profiles, the effects on the anisotropy is actually big ($\sim20-30\%$). The same consideration remains true even for more extreme cases. Second, the asymmetry between the two cases $s_3=\pm1$ is impressive, even for this moderate case: the two set of profiles are remarkably different, while the value of $\ai$, which gives us an idea of how far from gyrotropy ($\azi=0$) the system is, only reaches $\azi\simeq0.11$ ($\azi\simeq-0.11$) for the configuration with $s_3=+1$ ($s_3=-1$). The fundamental difference between the two configurations, $s_3=\pm1$, may then lead to very different dynamical evolution of the system and thus of the instability under study already in the linear phase~\cite{NaganoPSS1979,HubaGRL1996,HenriPOP2013}. 
In Fig.~\ref{fig:fig2} we show the same quantities as in Fig.~\ref{fig:fig1} for a different set of parameters: $\FF=\GG=\HH=1$, $u_0=3/2$, $L_u=3$, $B_0=\pm1$ ($s_3=\pm1$), $\betiprpz=\beteprpz=2$ and $\wgam=0$, which correspond to $\azi\simeq\pm0.25$.
\begin{figure}[!h]
  \begin{minipage}[!h]{0.49\textwidth}
  \flushleft\includegraphics[width=1.0\textwidth]{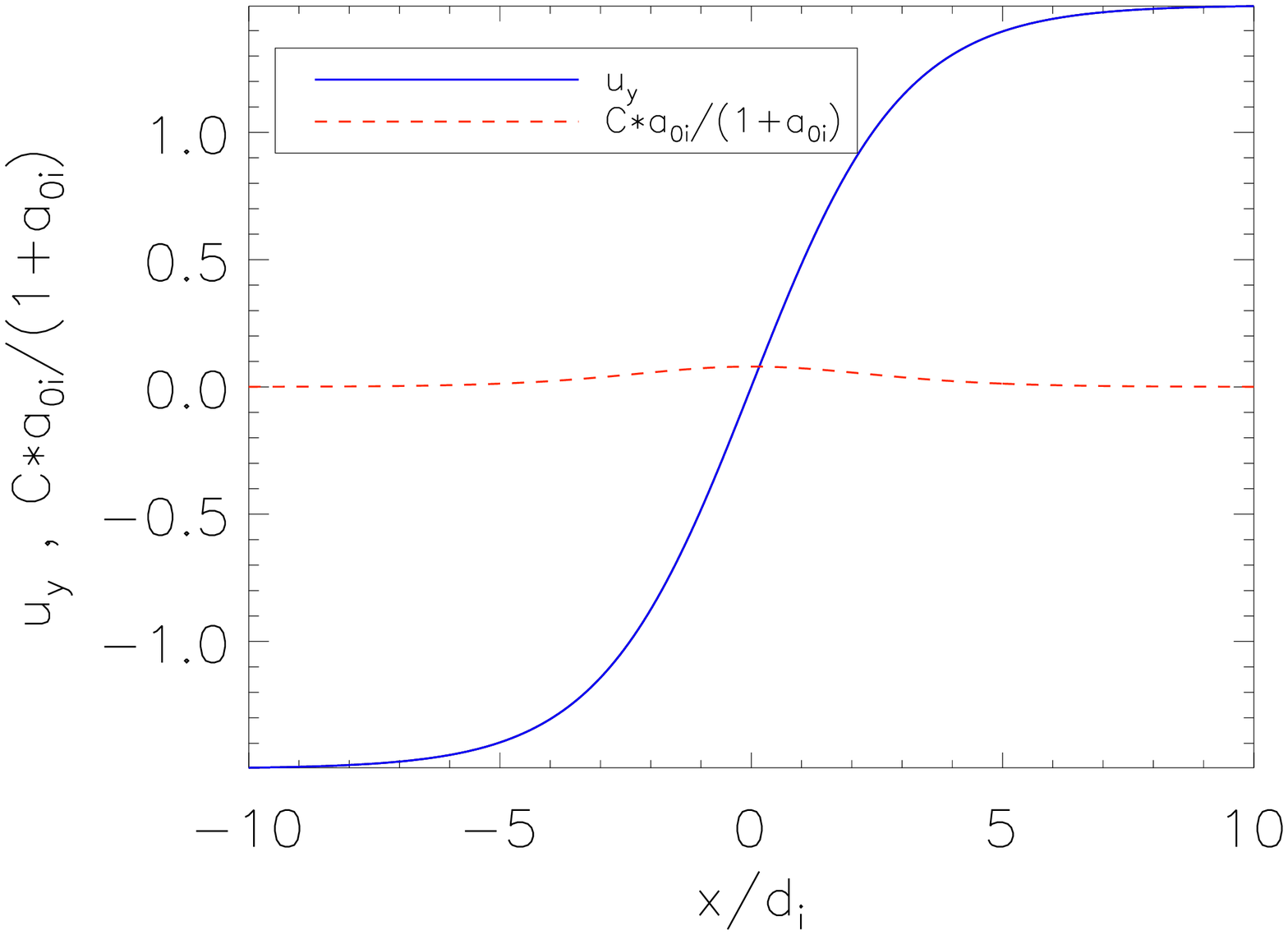}
  \end{minipage}
  \begin{minipage}[!h]{0.49\textwidth}
  \flushleft\includegraphics[width=1.0\textwidth]{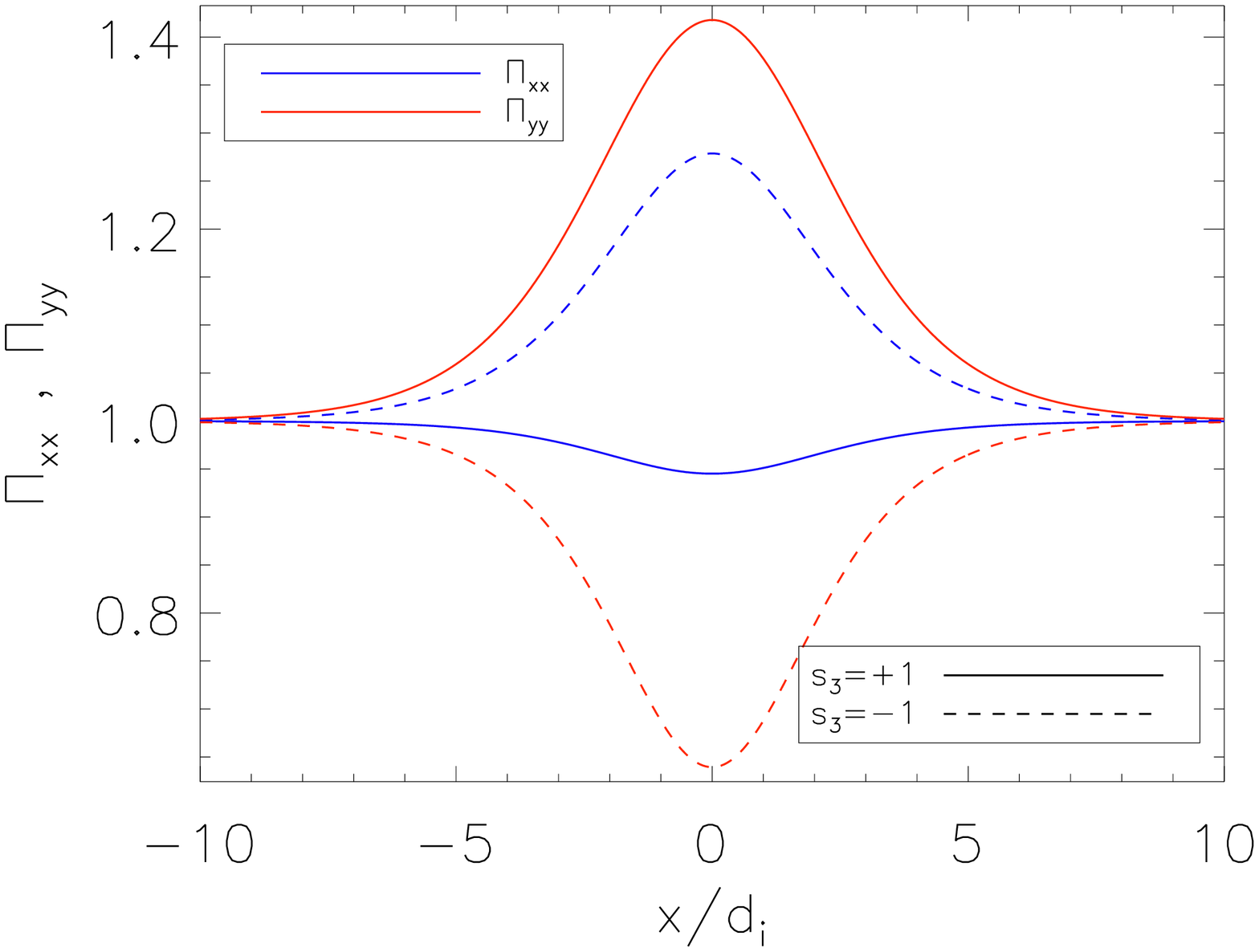}
  \end{minipage}
\caption{Left: velocity profile $\uiy(x)$ (blue solid line) and the function $\CC(x)\ai(x)/[1+\ai(x)]$ (red dashed line). Right: plot of the approximated equilibrium profiles $\Piixx(x)$ ($s_3=+1$: bottom blue solid line, $s_3=-1$: top blue dashed line) and $\Piiyy(x)$ ($s_3=+1$: top red solid line, $s_3=-1$: bottom red dashed line). Here, the parameters are: $\FF=\GG=\HH=1$, $u_0=3/2$, $L_u=3$, $B_0=\pm1$ ($s_3=\pm1$), $\betiprpz=\beteprpz=2$ and $\wgam=0$.}
\label{fig:fig2}
\end{figure} 

\section{Conclusions and discussion}\label{sec:end}

We have presented a study of the role of the pressure tensor in the presence of a sheared velocity field within a fluid plasma framework.
The heat fluxes are neglected. Solutions of the stationary pressure tensor equation are given for a simple, but commonly adopted configuration, and the properties of such equilibrium solutions are discussed. In particular, we have shown that, in addition to the well known parallel-perpendicular anisotropy ($p_\|\neq p_\perp$), the system is also anisotropic in the plane perpendicular to the magnetic field, i.e. $\Pixx\neq\Piyy\neq\Pizz$. The magnitude of the perpendicular anisotropy turns out to depend on the strength of the velocity shear and on its scale length of variation. Moreover, the system is strongly asymmetric with respect to the relative orientation of the background magnetic field and of the fluid vorticity, i.e. with respect to the sign of $\Omega_\uv\cdot\Bv$. These properties of the system are present even at the level of the equilibrium state representing the starting point for the study of shear-flow instabilities.

A method for deriving equilibrium profiles is presented and both numerical and approximated analytical solutions are provided for some representative cases. The profiles derived in the present paper are shown to be different with respect to the usual MHD or even CGL equilibria. In particular, they depend on the velocity shear and are asymmetric with respect to the sign of $\Omega_\uv\cdot\Bv$. These features, arising already at the level of the equilibrium configuration, turn out to be relevant when fluid models that retain the pressure tensor equation and/or kinetic models are adopted, as for the study of the KHI and the MRI.

Finally, despite the relative simplicity of the system configuration adopted, it seems plausible that our results can be used for the interpretation of satellite data where non-gyrotropic distribution functions are observed. This could be the case, for instance, of solar wind data, since, as pointed out by recent studies, one expects that the turbulence spontaneously generates local velocity shear flows.

\subsection*{Acknowledgments}
The research leading to these results received funding from the European Research Council under the European Unions Sevenths Framework Programme (FP7/2007-2013)/ERC Grant Agreement No. 277870. The research leading to these results has received funding from the European Commission's Seventh Framework Programme (FP7/2007-2013) under the grant agreement SWIFF (project No. 263340, www.swiff.eu).

\appendix{}

\section{Alternative analytical equilibria}\label{app:FLRequilibria_altern}

A relevant feature of Eq.~(\ref{eq:equil_cond_dimensionless}) is that it is general and versatile. Depending on the {\em physical} requirements, which then translate into mathematical relations between $f$, $g$ and $h$, one can compute very different equilibrium profiles. In the following, we give some examples.

\subsection{Preserving the total perpendicular plasma beta}

Requiring that the {\em total} perpendicular plasma beta $\betperp(x)$ remains unchanged in passing from MHD to full pressure tensor equilibria, the relation in Eq.~(\ref{eq:FLR_betaiperp_cond}) is substituted by
\begin{equation}\label{eq:FLR_betaperp_cond}
 h(x) = \xi_{\wgam}(\FF,f;x)f(x)\,,
\end{equation}
where the function
\begin{equation}\label{eq:xi_func_def}
 \xi_{\wgam}(\FF,f;x) \equiv 
 \frac{\betiprpz+\beteprpz\left(\FF(x)f(x)\right)^{\wgam}}{\betiprpz+\beteprpz\left(\FF(x)\right)^{\wgam}}\,,
\end{equation}
has been defined, such that it reduces to $\xi_0(\FF,f;x)=1$ for $\wgam=0$. For $\wgam=0$ we indeed recover Eq.~(\ref{eq:FLR_betaiperp_cond}) and thus the equilibrium condition in Eq.~(\ref{eq:FLR_equil_cond}) and its solution. For $\wgam\neq0$, the equilibrium condition reads
\[
 \Bigg\{\xi_{\wgam}(\FF,f;x)\ +\ 
 \wbetiprpz\left[\frac{1-\xi_{\wgam}(\FF,f;x)\left(1+\azi(x)\right)}{1+\azi(x)}\right]\FF(x)\ +
\]
\begin{equation}\label{eq:FLR_equil_cond_2}
 \wbeteprpz\left(\FF(x)\right)^{1+\wgam}\left(f^{\wgam}(x)-\xi_{\wgam}(\FF,f;x)\right)\Bigg\}f(x)\ =\ 1\,,
\end{equation}
which, Taylor expanding $f^{\wgam}$ and $\xi_{\wgam}$ as in Sec.\ref{subsec:FLRequilibria_approx} and after some algebra, admits the solution
\begin{subequations}\label{eq:FLR_equil_conv_2}
\begin{equation}
 f(x) = \left[1-\CC(x)\frac{\ai(x)}{1+\ai(x)}\right]^{-1}\,,
\end{equation}
\begin{equation}
 \CC(x) = \left[1+\wgam\wbeteprpz\wFFg(x)\right]^{-1}\CC_0(x)\,.
\end{equation}
\begin{equation}
 \wFFg(x) \equiv \frac{\left(\FF(x)\right)^{\wgam}}{\wbetiprpz+\wbeteprpz\left(\FF(x)\right)^{\wgam}}\,.
\end{equation}
\end{subequations}

\subsection{Preserving the magnetic field configuration}

We may want to fix the magnetic field configuration: this is equivalent to requiring that the magnetic field profile remains unchanged with respect to the MHD profile, i.e. 
\begin{equation}\label{eq:FLR_B-unchanged_cond}
 h(x) = 1\quad,\quad
 \frac{\HH(x)}{1+\betprpz} = 1 - \left[\wbetiprpz + \wbeteprpz\left(\FF(x)\right)^{\wgam}\right]\FF(x)\,,
\end{equation}
and thus we need to solve the following equilibrium condition:
\begin{equation}\label{eq:FLR_equil_cond_3}
 \left[\wbetiprpz\left(1-\frac{\ai(x)}{1+\ai(x)}\right) +
 \wbeteprpz\FF^{\wgam}(x)f^{\wgam}(x)\right]f(x)\ =\ \wbetiprpz+\wbeteprpz\FF^{\wgam}(x)\,,
\end{equation}
which, considering the case $\wgam=0$ for simplicity, has the solution 
\begin{equation}\label{eq:FLR_equil_0_3}
 f_0(x) = \frac{1+\tauprp}{1+\tauprp - \frac{\azi(x)}{\sqrt{\HH(x)}+\azi(x)}}\,,
\end{equation}
where we have defined the perpendicular temperature ratio $\tauprp\equiv\Teperpz/\Tiperpz$ for brevity and now the function $\ai(x)$ is computed with the actual local magnetic field, without approximations (i.e., $\ai(x)\equiv\azi(x)$).


\end{document}